\documentclass[preprint,12pt]{elsarticle}

\usepackage{amssymb}
\usepackage{amsmath}
\usepackage{mathrsfs}
\usepackage{mathtools}

\usepackage{algorithm}
\usepackage{algpseudocode}

\usepackage{tabularx}
\usepackage{booktabs}

\usepackage{verbatim}

\usepackage{gensymb}

\usepackage[printonlyused]{acronym}
\usepackage[hang,flushmargin]{footmisc}
\newcommand\extrafootertext[1]{%
    \bgroup
    \renewcommand\thefootnote{\fnsymbol{footnote}}%
    \renewcommand\thempfootnote{\fnsymbol{mpfootnote}}%
    \footnotetext[0]{#1}%
    \egroup
}

\usepackage{hyperref}

\usepackage{subfig}

\usepackage{comment}

\biboptions{numbers,sort&compress}

\journal{Energy}

\begin{document}

\begin{frontmatter}



\title{Enabling Predictive Maintenance in District Heating Substations: A Labelled Dataset and Fault Detection Evaluation Framework based on Service Data}

%% Author name
\author[fhiee]{Cyriana M.A. Roelofs\corref{cor1}} 
\ead{cyriana.roelofs@iee.fraunhofer.de}
\author[fhiee]{Edison Guevara Bastidas}
\author[fhiee]{Thomas Hugo}
\author[fhiee]{Stefan Faulstich}
\author[fhiee]{Anna Cadenbach}
\cortext[cor1]{Corresponding author}

\affiliation[fhiee]{
organization={Fraunhofer IEE},
addressline={Joseph-Beuys-Straße 8, 34117 Kassel}, 
country={Germany}
}

%% Abstract
\begin{abstract}
%% Text of abstract - max 250 words
Early detection of faults in district heating substations is imperative to reduce return temperatures and enhance efficiency. However, progress in this domain has been hindered by the limited availability of public, labelled datasets. We present an open-source framework combining a service-report-validated public dataset, an evaluation method based on accuracy, reliability, and earliness, and baseline results implemented with EnergyFaultDetector, an open-source Python framework developed for automated anomaly detection in operational data from energy systems.

The dataset contains time series of operational data from 93 substations across two manufacturers, annotated with a list of disturbances due to faults and maintenance actions, a set of normal-event examples and detailed fault metadata. We evaluate EnergyFaultDetector models, adapted and configured for district heating substations, using three metrics: accuracy for recognising normal behaviour, an eventwise F-score for reliable fault detection with few false alarms, and earliness for early detection. The framework also supports root cause analysis using ARCANA,  a feature‑attribution method for autoencoders. We demonstrate three use cases to assist operators in interpreting anomalies and identifying underlying faults. The models achieve high normal‑behaviour accuracy (0.98) and eventwise $F_{0.5}$ of 0.83 and could detect 60\% of the faults in the dataset before the customer reported a problem, with an average lead time of 3 to 5 days. 

Integrating an open dataset, metrics, open-source code, and baselines establishes a reproducible, fault-centric benchmark with operationally meaningful evaluation, enabling consistent comparison and development of early fault detection and diagnosis methods for district heating substations.

\end{abstract}

% Graphical abstract not needed for Energy
\begin{graphicalabstract}
\includegraphics[width=\textwidth]{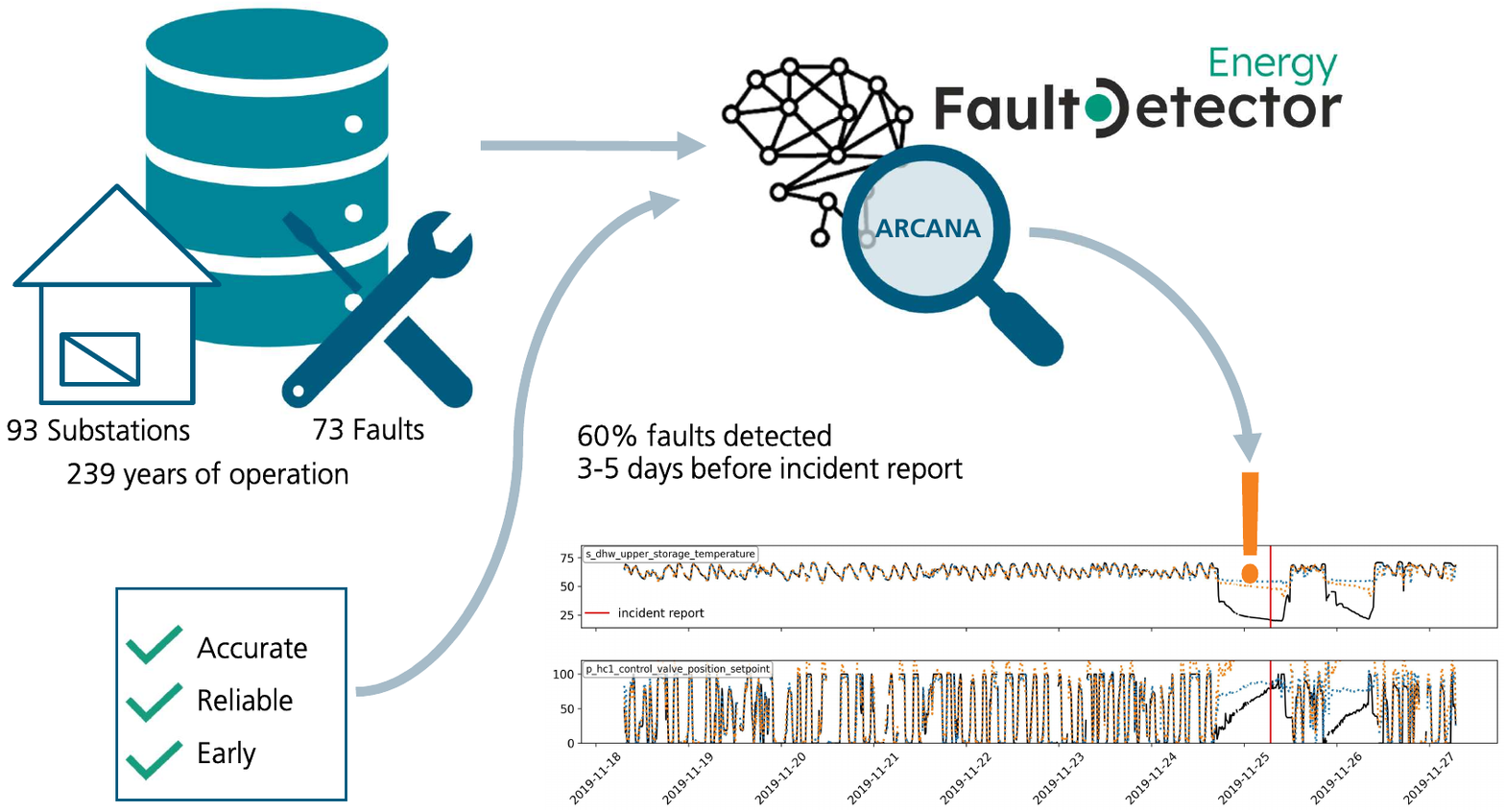}
\end{graphicalabstract}

%Not part of editorial consideration and aren't required until the final files stage
%Only required for full research articles
%Must be provided as a Word document— select "Highlights" from the drop-down list when uploading files
%Each Highlight can be no more than 85 characters, including spaces
%No jargon, acronyms, or abbreviations: aim for a general audience and use keywords
%Consider the reader - Highlights are the first thing they'll see
\begin{highlights}
\item Public, service-report–validated labelled dataset of 93 district heating substations.
\item Fault labels, metadata, and open‑source baseline enable benchmarking.
\item Practical scoring method rewards early and reliable fault detection.
\item Open‑source baseline and root-cause-analysis with EnergyFaultDetector.
\item The models could detect 60\% of the faults before a customer reported the problem.
\end{highlights}

%% Keywords

\end{frontmatter}



\section*{Nomenclature}
\begin{acronym}
\acro{dh}[DH]{district heating}
\acro{dhs}[DHS]{district heating substation}
\acro{ae}[AE]{autoencoder}
\acro{nbm}[NBM]{normal behaviour model}
\acro{scada}[SCADA]{Supervisory Control and Data Acquisition}
\acro{mse}[MSE]{mean square error}
\acro{rmse}[RMSE]{root mean square error}
\acro{re}[RE]{reconstruction error}
\acro{om}[O\&M]{Operation \& Maintenance}
\acro{ml}[ML]{machine learning}
\acro{dl}[DL]{deep learning}
\acro{fdd}[FDD]{fault detection and diagnosis}
\acro{dhw}[DHW]{domestic hot water}
\acro{sh}[SH]{space heating}
\acrodef{mpn}[MPN]{Maintenance Priority Number}
\end{acronym}

% FLA: max 8000 words
    \section{Introduction}\label{introduction}
% Early fault detection --> predictive maintenance --> measueres or actionable information to reduce on-call maintenance interventions (maintenance outside regular working hours) --> lack of technicians --> thousends of stations

District heating plays a crucial role in the large-scale integration of renewable energy for environmentally friendly heat supply \cite{LUN2018}. The increase of efficiency in district heating networks and the decrease of distribution temperatures \cite{LUN2014} is essential to unlock that potential. 

Detecting faults in district heating substations is key for both technical and operational reasons. On the one hand, identifying and correcting faults that lead to elevated return temperatures is essential for enabling lower supply temperatures across the network, which in turn reduces distribution flows and enhances the overall energy efficiency of the system \cite{gadd_achieving_2014}. On the other hand, as the number of substations continues to grow \cite{AGORA2024}, and due to the lack of appropriate monitoring tools and service personnel, utilities face increasing pressure to operate and maintain these assets efficiently while ensuring a secure and reliable heat supply. 

Digitalisation measures for the collection of customers' operating data and novel machine learning methods enable the development of data-driven solutions for anomaly detection \cite{neumayer_fault_2023}. These developments together with efforts in the digitalisation and structuring of maintenance information in substations \cite{mansson_taxonomy_2021}, lay the basis for the development of methods for early fault detection and diagnosis, enabling utilities to optimise their maintenance strategies and implement predictive maintenance in their O\&M processes.

Many different failure modes can occur in substations. M\r{a}nsson et al. \cite{mansson_faults_2019} identify six fault categories based on surveys of practitioners in Sweden: heat exchangers, control systems and controllers, actuators, control valves, the customer’s internal heating system and leakages. The study reports that leakages are the most common category of faults, followed by faults in the customers’ internal heating systems. Gadd and Werner \cite{gadd_fault_2015} focus on the effects on the efficiency and divide faults into three groups: unsuitable heat load pattern, low average annual temperature difference, and poor substation control, with 70\% of the analysed Swedish substations showing low annual temperature difference. Østergaard et al. \cite{oestergaard_2022} and Leoni et al. \cite{LEONI_2020} similarly report frequent issues with heating system components, control valves and actuators, incorrect setpoints, and leakages.

While the causes of substation faults are manifold, not all failure modes are relevant to early fault detection methods and, consequently, predictive maintenance. Some have clear, monitorable signatures in operational data, while others require additional sensors or organisational measures on the building side \cite{guevara_bastidas_prioritisation_2025}. Predictive maintenance depends on early, reliable, and interpretable detection. \ac{fdd} systems provide these signals from operational data, ideally with enough lead time to act.

However, progress in intelligent \ac{fdd} for \ac{dhs} is hindered by the lack of public, labelled, real-world datasets and consistent evaluation metrics \cite{van_dreven_intelligent_2023}. Most studies rely on a single, generally unlabelled dataset provided by a cooperating utility company \cite{neumayer_fault_2023}, while public datasets are often simulation-based to protect customer privacy. As each study uses a different dataset and evaluation metric, it is difficult to compare the results of available methods in the literature.

This study provides a framework for the evaluation of early fault detection methods, enabling the comparison between approaches and, consequently, promoting the development of \ac{fdd} techniques for \ac{dhs}. The contribution and novelty of this study comprise three elements: 
\begin{enumerate}
    \item The main contribution of this paper is the publication of a labelled dataset\footnote{The PreDist dataset v1 can be found on Zenodo: \url{https://doi.org/10.5281/zenodo.17522255 } under the CC BY 4.0 license.} containing time series of 93 \ac{dhs} belonging to one operator. This includes 10-minute operational data, timestamps of disturbances due to maintenance tasks,  incident reports from customers, and derived fault labels. To the best of our knowledge, this is the first publicly available, service-report–validated labelled dataset for fault detection in \ac{dhs}.
    \item Additionally, we provide an open-source fault detection baseline for \ac{dhs} using the `EnergyFaultDetector'\footnote{Github: \url{https://github.com/AEFDI/EnergyFaultDetector/tree/v0.3.0}}, an autoencoder-based fault detection Python framework, originally developed for early \ac{fdd} in wind turbines \cite{roelofs_autoencoder-based_2021}, and here adapted and validated for \acp{dhs}.
    \item We define an evaluation protocol for early fault detection in DHS based on three metrics inspired by the CARE score \cite{guck_care_2024}: accuracy, reliability and earliness. We adapt the earliness score to work with datasets where the exact fault onset is unknown by comparing the detection timestamp with an actionable window before the incident report. Together with the dataset and baseline, this yields a reproducible, fault-centric benchmark for early fault detection in DHS. 
\end{enumerate}

%The paper is structured as follows: ... 

\section{Related work}\label{related work}

\subsection{Fault detection in district heating substations}
A thorough overview of methodologies and datasets used is provided by Neumayer et al. \cite{neumayer_fault_2023}, while van Dreven et al. discuss recent trends and practical challenges for intelligent \ac{fdd} in district heating \cite{van_dreven_intelligent_2023}.

Statistical and rule‑based methods are a common baseline because they are simple, interpretable and require little labelled data: manual visual inspection \cite{gadd_fault_2015}, fixed or adaptive thresholds (e.g. three‑sigma) \cite{gadd_achieving_2014}, piecewise linear regression \cite{gadd_fault_2015,mansson_automated_2018, theusch_fault_2021} and simple physical models based on performance-signature metrics (such as $\Delta$T or excess flow) \cite{leiria_towards_2023} are typically used to identify sub-optimally performing substations. These methods work well for broad screening and operator workflows but struggle with seasonality, heterogeneous user behaviour and coarse sampling (many studies use hourly billing data), which leads to high false‑alarm rates or the need for substantial manual follow‑up \cite{neumayer_fault_2023}.

From roughly 2018 onwards there has been a clear shift towards \ac{ml} approaches as more metering data became available \cite{van_dreven_intelligent_2023}. However, because labelled faults are scarce most work remains unsupervised or semi‑supervised. Typical unsupervised and semi-supervised techniques include clustering and pattern mining \cite{calikus_data-driven_2019, abghari_higher_2019}, regression with residual analysis \cite{gadd_achieving_2014, mansson_automated_2018, theusch_fault_2021} and outlier detection algorithms such isolation forests \cite{farouq_large-scale_2020}.

Deep learning methods, such as \acf{ae}, have recently been investigated for reconstruction‑based anomaly detection and latent‑space feature learning in substations \cite{zhang_anomaly_2020, choi_autoencoder-driven_2021}. These models can capture complex, non‑linear multivariate temporal signatures and improve sensitivity to subtle faults, but they typically require more data and explainability measures due to their blackbox nature.

\subsection{Public datasets}
As mentioned in the introduction, many publicly available datasets are simulation-based. One example is presented by Vallée et al. \cite{vallee_generation_2023} and is partly available on Kaggle \cite{vallee_kaggle_2024}. It was created as a reference dataset for fault detection in \ac{dh} systems, covering production, distribution, substations and storage with diverse fault profiles and reproducible generation procedures. The authors benchmark several classical \ac{ml} models and report strong performance on global-efficiency faults but difficulty detecting thermal-loss faults, and they show encouraging preliminary transfer results to limited real-world data.

Van Dreven et al. \cite{van_dreven_systematic_2024} present a systematic laboratory emulation of a generic district-heating substation (connected to a climate chamber) to generate labelled time-series data for five fault types, validated against two real substations. Using a fault-detection pipeline (isolation forest and one‑class SVM for detection, random forest and SVM for diagnosis), they show that higher sampling frequencies improve performance (5‑min optimal for detection, 1‑min for diagnosis), while noting limitations in emulating long-term and user-driven behaviours.

Next to simulation-based and laboratory-emulation-based datasets, recent work addresses the lack of shareable labelled data by generating synthetic data based on a large, non-public, fully reviewed dataset (ILSE) \cite{stecher_neural_2025}. The authors explore three \ac{ml}-based synthetic data generation approaches: time-series forecasting (FCNN/LSTM), GAN-based generation (TimeGAN), and fault-signature transfer with Transformer models. However, synthetic fault data were not yet reliable enough to train fault detectors, and fault-signature transfer struggled due to high fault-variance and limited samples. This highlights the current limitations of synthetic data for \ac{fdd} in \ac{dhs}, and it further motivates public, service-validated labelled datasets and reproducible benchmarks.

To the best of the authors' knowledge, there is currently no publicly available, real-world annotated dataset for fault detection in district heating substations.

\section{PreDist Dataset}\label{data}
The PreDist dataset consists of time series containing raw operational data from 93 \ac{dhs} with control units by two manufacturers, \emph{M1} and \emph{M2}, and operated by a single utility. In addition, the time series are annotated with maintenance and incident reports.

All \ac{dhs} present in the dataset are indirectly connected to the district heating network and are of one of the following configuration types:

\begin{itemize}    
    \item \texttt{SH + DHW}: one space heating circuit and one \ac{dhw} circuit with storage tank
    \item \texttt{SH}: only one space heating circuit 
    \item \texttt{SH + DHW with sub-circuits}: one space heating circuit with additional sub-circuits and one \ac{dhw} circuit with storage tank
    \item \texttt{SH with buffer tank}: one space heating circuit with a buffer tank
    \item \texttt{SH with sub-circuits}: one space heating circuit with additional sub-circuits
\end{itemize}

\begin{figure} 
    \centering
    \includegraphics[width=.8\linewidth]{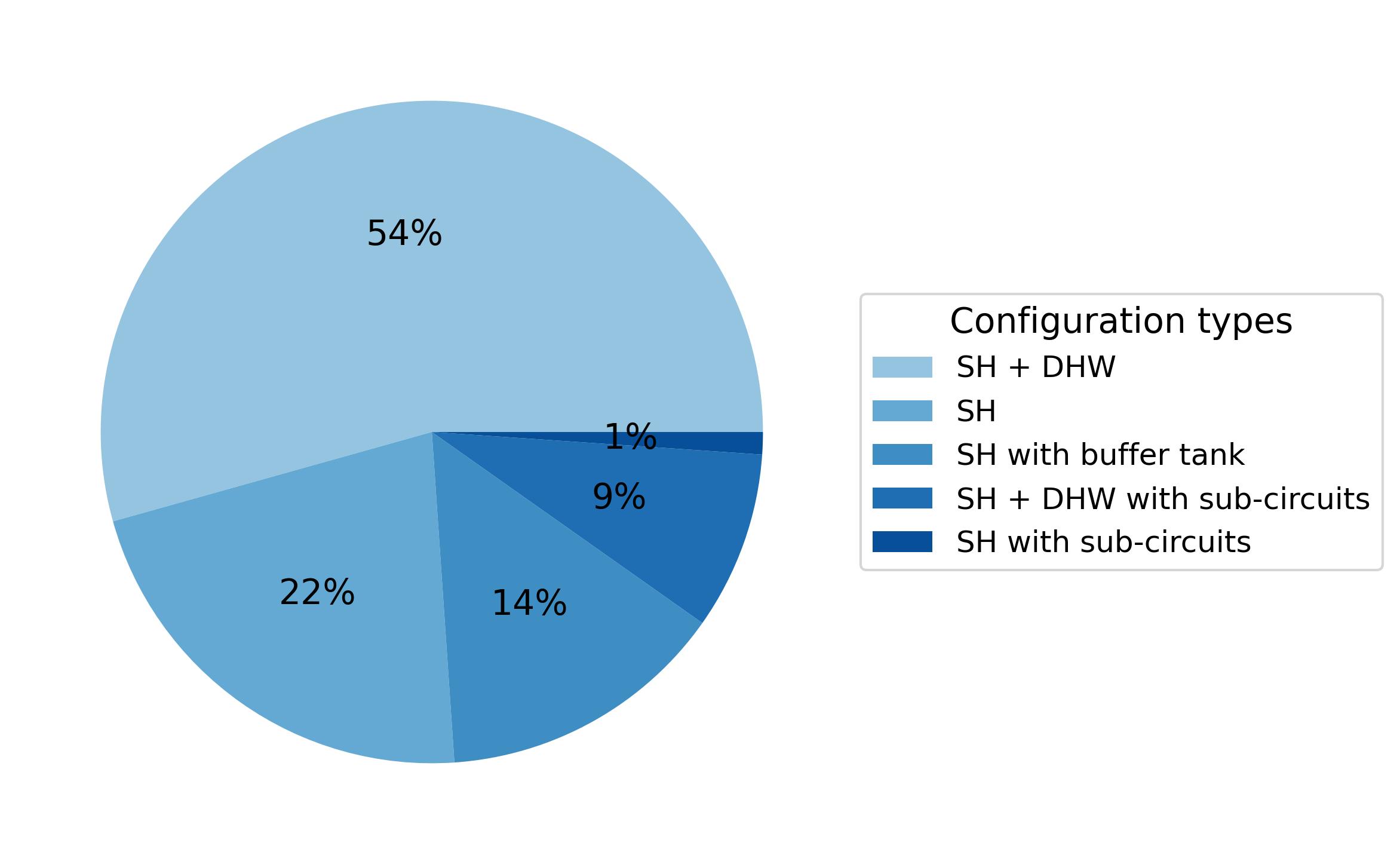}
    \caption{Overall share of configuration types in the dataset, covering both manufacturers.}
    \label{fig:share_of_conf_all_stations}
\end{figure}

The main configurations present in the dataset are \texttt{SH + DHW} and \texttt{SH} representing together 76\% of all \ac{dhs}. Figure \ref{fig:share_of_conf_all_stations} shows the share of configuration types across all substations. While we cannot assess whether this configuration mix is representative for other district heating networks due to a lack of public statistics for Germany or other countries, it is representative for the operator who provided the data.

\begin{figure}
    \centering
    \includegraphics[width=.9\linewidth]{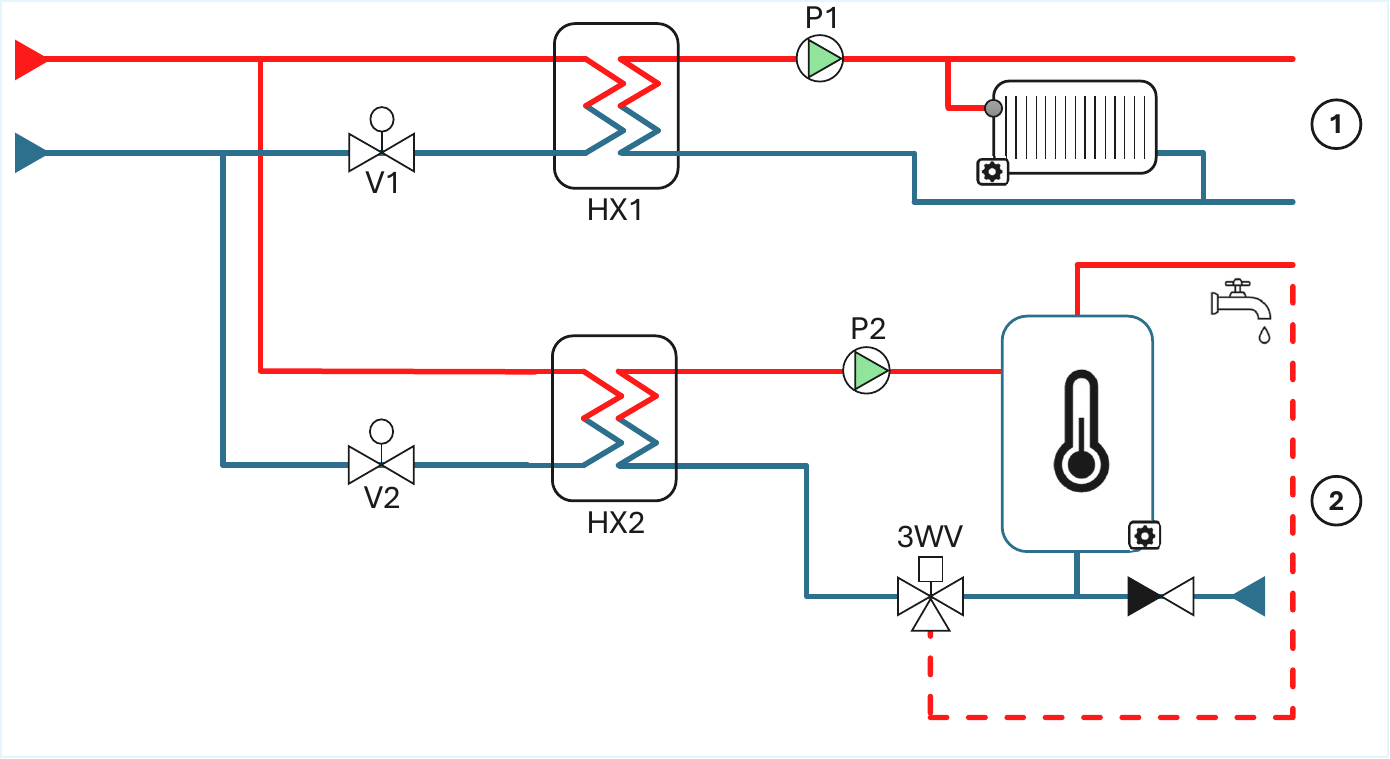}
    \caption{Schematic representation of the \ac{dhs} configuration \texttt{SH + DHW}, which is most common in the PreDist dataset. The figure shows the space heating circuit (1) and the \ac{dhw} circuit (2), both indirectly connected to the district heating network via a heat exchanger, as well as their main components. Circuit (1) comprises control valve V1 and circulation pump P1; circuit (2) comprises control valve V2, charging pump P2, a storage tank, and a three-way valve (3WV). The configuration \texttt{SH} consists only of circuit (1).}
    \label{fig:schematic_sh_dhw}
\end{figure}

A schematic representation of configurations \texttt{SH + DHW} and \texttt{SH} is shown in Figure~\ref{fig:schematic_sh_dhw}. Both configurations exhibit one heating circuit indirectly connected to the \ac{dh} network by means of a heat exchanger. The main components of the circuit are a motorised control valve on the primary side and a circulation pump on the secondary side. Configuration \texttt{SH + DHW} has additionally a \ac{dhw} circuit indirectly connected to the \ac{dh} network by means of a separate heat exchanger. The main components of the circuit are a motorised control valve on the primary side, and on the secondary side a 3-way-valve, a storage tank and a charging pump.

Typical operating ranges in the PreDist dataset are as follows: primary supply temperatures mostly between 80$^{\degree}$C and 110$^{\degree}$C, with return temperatures typically 20 to 40K lower. The secondary space-heating supply temperatures vary between roughly 30$^{\degree}$C and 70$^{\degree}$C depending on outdoor conditions and control unit settings.

The data are anonymised by removing personally identifiable information, such as names, phone numbers, and addresses, from the fault descriptions. Problem descriptions and maintenance measures were combined and summarised, and feature names in the time series were standardised for both manufacturers. All timestamps were shifted by a single, undisclosed constant offset of several years, applied consistently across all substations and data sources to preserve relative timing. The shift preserves time-of-day, day-of-week and seasonal patterns.

\subsection{Operational data}\label{operational_data}
The time series have a 10-minute resolution and contain instantaneous measurements (i.e., not 10-minute aggregates) such as primary and secondary supply temperatures, setpoints (e.g. for the secondary supply temperature), and status values (e.g., circulation pump on/off). Depending on the configuration of the substation, e.g. whether a separate \ac{dhw} circuit is present, the number of features varies across substations. For both \emph{M1} and \emph{M2} datasets, a list of features and their units of measurement is provided as CSV files with the dataset on Zenodo.

The lengths of the time series also vary across substations: the number of substations connected to the network increased over time, so some series span multiple years while others span less than one year. An overview of the dataset's statistics is shown in Table~\ref{tab:dataset_overview}, showing the number of substations, average time series length, total time series length, number of features, average number of annual incident reports and maintenance tasks and the average completeness of the time series. Completeness is defined as the percentage of available data points with respect to the expected number of samples at a 10-minute sampling rate, taking all features from both the energy meter and the control unit into account. If only the control unit or only the meter data are present, this is counted as an incomplete timestamp. The time series are the raw records provided by the utility and they may contain implausible values, missing values, and data gaps that must be accounted for when analysing the data and developing \ac{ml} models based on this dataset.

\begin{table}[ht]
\centering
%\resizebox{\textwidth}{!}{%
\begin{tabular}{@{}llll@{}}
\toprule                                        & \textbf{\emph{M1}} & \textbf{\emph{M2}} & \textbf{Total} \\ \midrule
\# substations                                  & 35   & 58  & 93   \\ 
Avg time series length (years)                  & 4.1  & 1.6 & 2.6  \\ 
Total time series length (years)                & 144  & 95  & 239  \\ 
\# Substations with time series $\ge$ 1 year     & 30   & 19  & 49   \\ %& 5    & 39  & 44   \\ 
\# features                                     & 10 -- 24   & 13 -- 41  & 10 -- 41   \\ 
Avg \# incident reports / year                  & 0.74  & 2.0 & 1.5  \\ 
\begin{tabular}[c]{@{}l@{}}Avg \# maintenance tasks  / year \\\end{tabular} & 0.85    & 1.4     & 1.2            \\ 
Avg completeness (\%)                           & 82   & 97  & 91  
\end{tabular}%
%}
\caption{Overview of the dataset. Averages are calculated across all substations. \emph{M1} and \emph{M2} refer to the two different control-unit manufacturers.}
\label{tab:dataset_overview}
\end{table}

\subsection{Maintenance and incident reports}\label{maintenance_data}
In addition to the time series, we provide a list of disturbances and a separate list of incident reports. The disturbances list records timestamps for incident reports, labelled as `fault', and for maintenance actions (preventive or corrective), labelled as `task'.
Disturbances can be used to select time ranges without faults or maintenance interventions that represent expected normal behaviour.

The incident reports list comprises reports with at least two weeks of operational data preceding the report, in which a fault was confirmed and subsequently corrected. Faults range from incorrect control unit or component settings to broken components that required replacement. Note that these faults are based on customer-reported incidents (i.e., issues affecting comfort). Faults relating solely to reduced efficiency may not be labelled in this dataset, as they may go unnoticed by customers.

\begin{figure}
    \centering
    \includegraphics[width=0.6\linewidth]{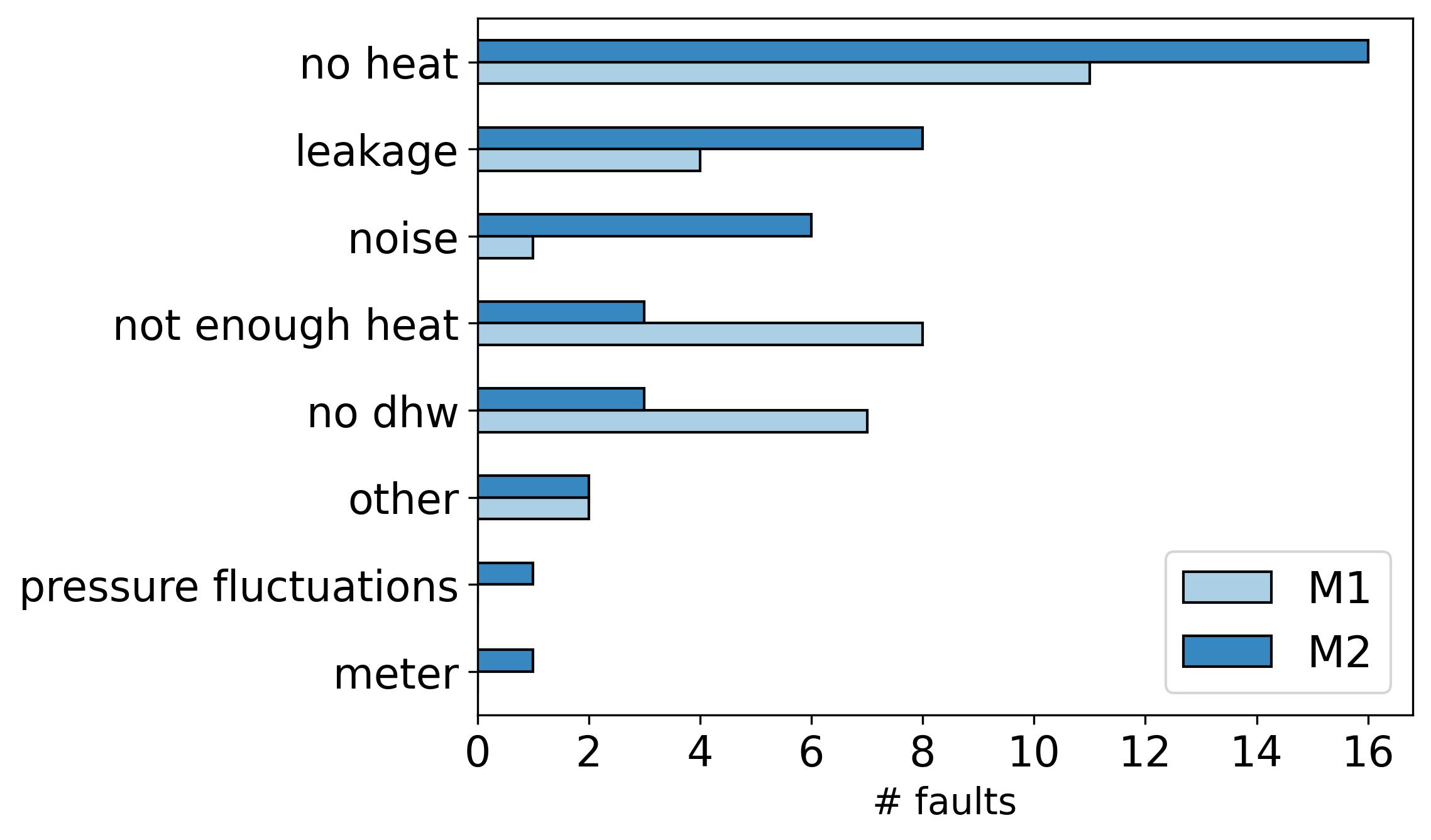}
    \caption{Number of faults per problem category and manufacturer (designated by \emph{M1} and \emph{M2}) in the dataset.}
    \label{fig:faults_per_problem_category}
\end{figure}

For each fault, a problem category (e.g. no heat) and, if known, a short description of the underlying cause and solution are provided. An overview of the number of faults per category is shown in Figure~\ref{fig:faults_per_problem_category}. In addition, we mark reports for which early fault detection is not possible from a data perspective. Examples include reports lacking normal-behaviour data prior to the incident and repeat incident reports, for which only the first occurrence can be considered. Because these reports may still be valuable (e.g., for testing fault diagnosis tools), we did not remove them from the final report list. 

\begin{figure}
    \centering
    \includegraphics[width=.9\linewidth]{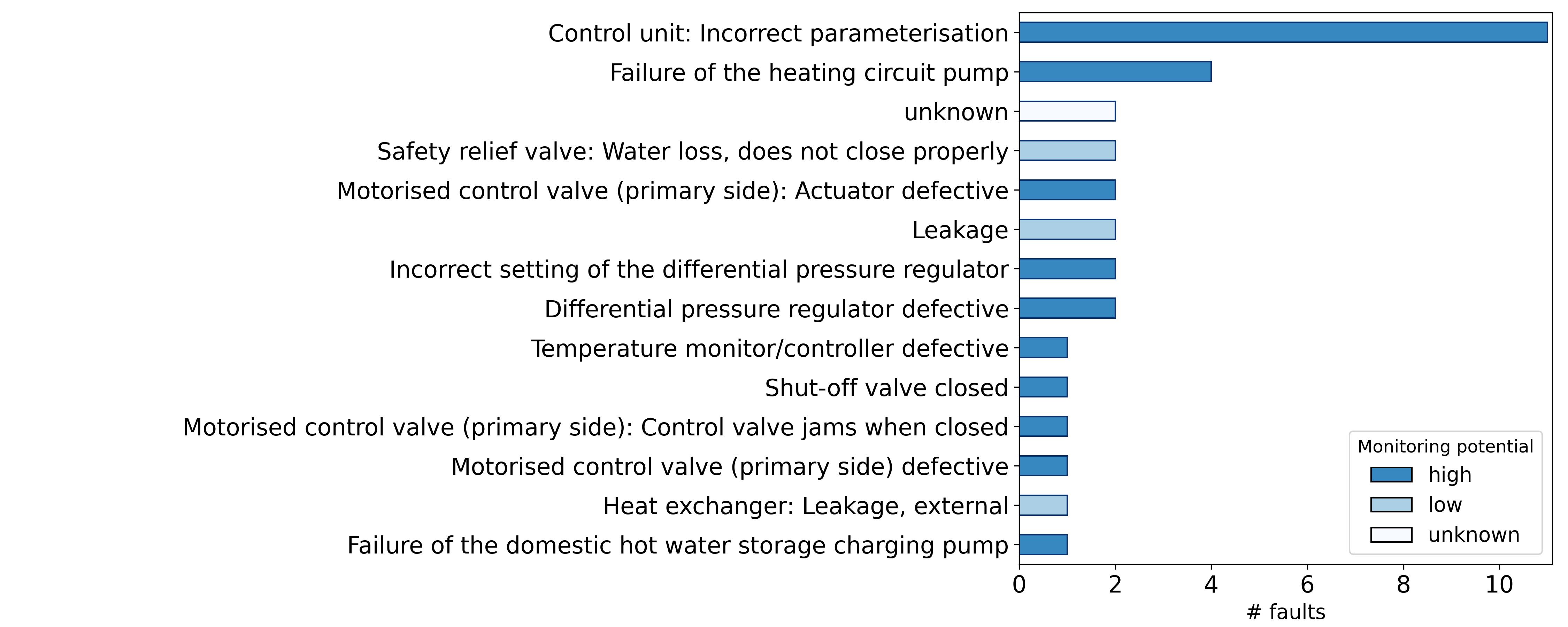}
    \caption{Fault label counts for incident reports of manufacturer \emph{M1}, showing their categorisation in monitoring potential.}
    \label{fig:fault_labels_m1}
\end{figure}

\begin{figure}
    \centering
    \includegraphics[width=.9\linewidth]{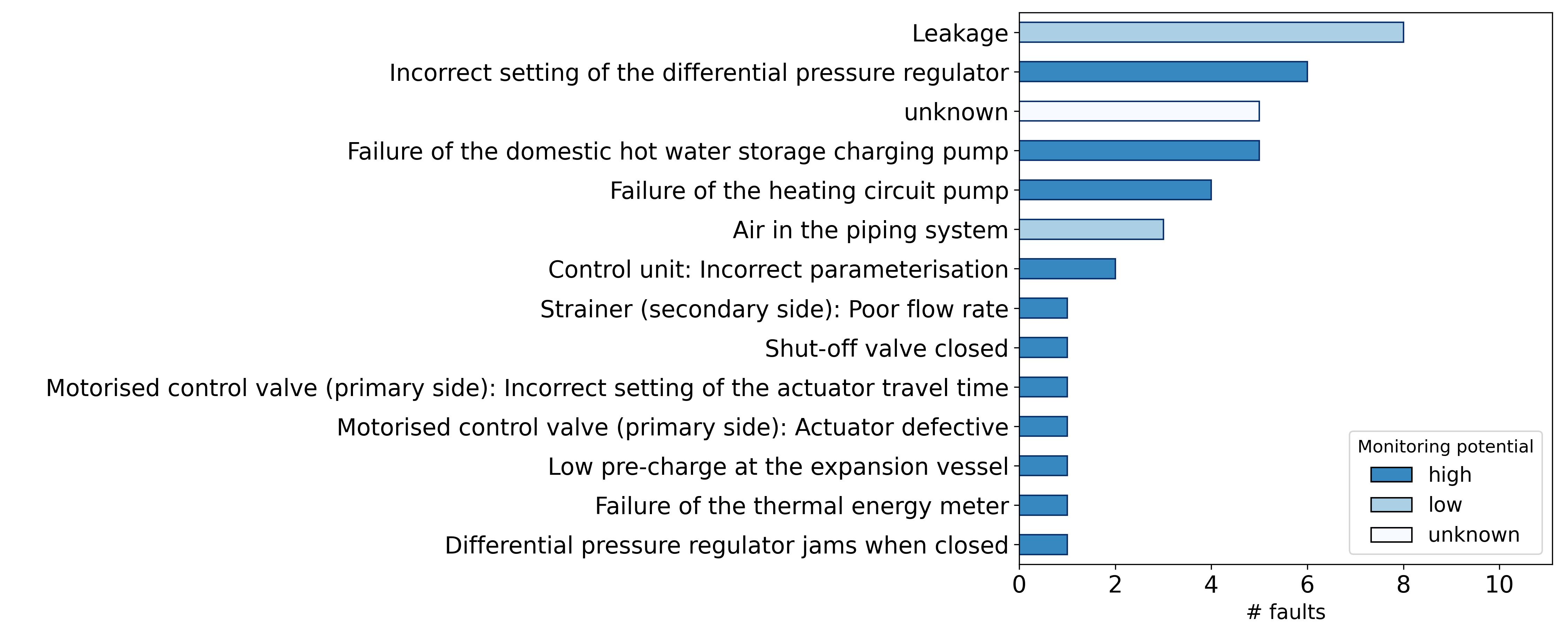}
    \caption{Fault label counts for incident reports of manufacturer \emph{M2}, showing their categorisation in monitoring potential.}
    \label{fig:fault_labels_m2}
\end{figure}

Each fault has been assigned, where possible, a fault label and a monitoring potential, aligned with \cite{guevara_bastidas_prioritisation_2025}. Following \cite{guevara_bastidas_prioritisation_2025}, the monitoring potential is rated on a 1–5 scale that reflects whether detection is possible before or only after fault manifestation and whether existing instrumentation suffices or additional effort is required. Faults that can be detected before the failure occurs have a higher rating than faults that can only be detected after the failure. Additionally, faults that can be detected with existing sensor measurements have a higher rating. For analysis, we bin the numeric rating into two classes: “high” (rating $\geq 2.5$) and “low” (rating $< 2.5$), reflecting suitability for data‑driven detection with available measurements. If the fault could not be assigned a fault label, the label and monitoring potential are set to 'unknown'. Faults labeled `unknown' were confirmed and corrected, but the underlying technical cause could not be clearly identified from the available information. These faults are included as positive events for fault detection evaluation. Ratings were assigned based on incident descriptions, maintenance logs, substation configuration, and visual inspection of time series where necessary. Ambiguous cases were given a more generic fault label and an averaged rating across plausible failure modes. Figures \ref{fig:fault_labels_m1} and \ref{fig:fault_labels_m2} show the overview of fault label occurrences, as well as their monitoring potential, for substations of manufacturers \emph{M1} and \emph{M2} respectively.


The categorisation of incident reports into fault labels and monitoring potential was sometimes ambiguous and required assumptions. These labels and ratings support evaluation and interpretation and should be treated as expert estimates rather than ground truth. Pump‑related issues, whether changes in settings or defects, were assigned one monitoring‑potential rating in line with \cite{guevara_bastidas_prioritisation_2025}. The monitoring potential of component and non‑specific leakages were set to 1.75, the mean rating for external leakages. When reports lacked detail, we used a general label and the average rating across plausible faults. A wrong adjustment of the needle throttle valve in the differential pressure regulator was mapped to ``Incorrect setting of the differential pressure regulator,” which may overestimate monitoring potential if the issue was noise or minor oscillations. Some reports could not be labelled due to insufficient information.

\begin{figure}
    \centering
    \includegraphics[width=\linewidth]{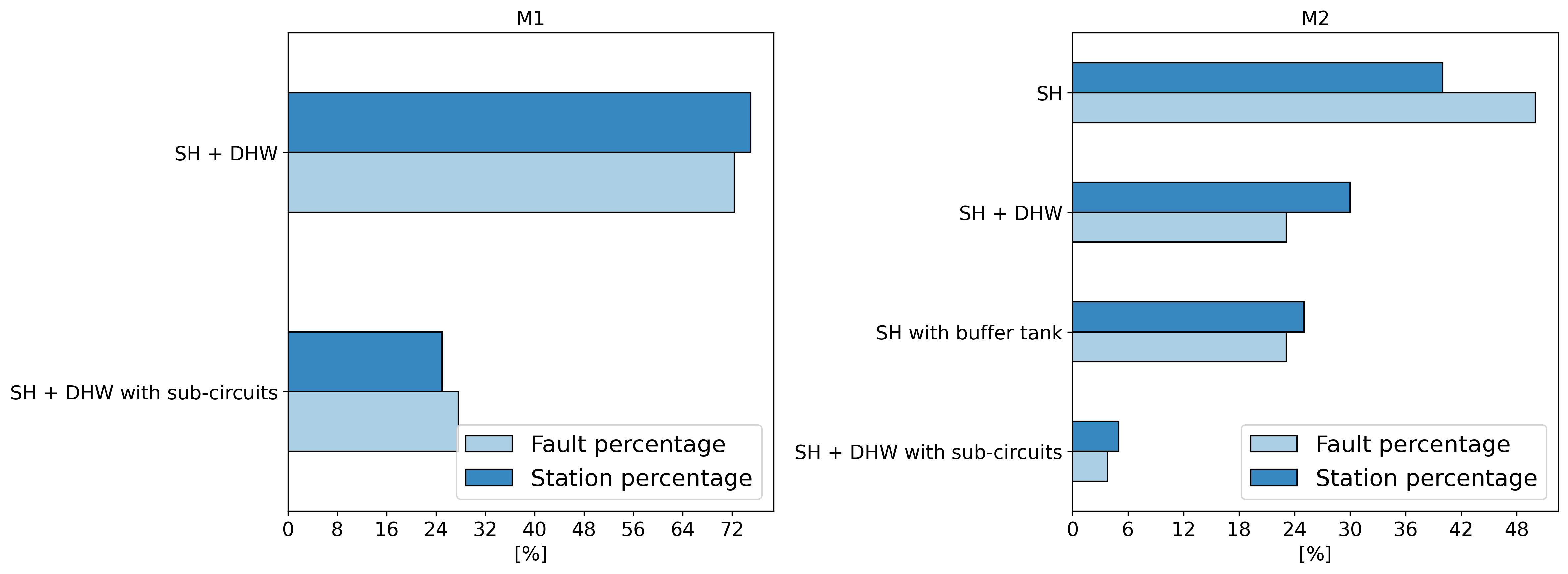}
    \caption{Share of configuration types across faults for manufacturers \emph{M1} and \emph{M2}.}
    \label{fig:share_of_faults_conf}
\end{figure}

An important aspect of the quality of the dataset is that the distribution of faults is well balanced across all \ac{dhs} configuration types present in the dataset. As can be seen in figure \ref{fig:share_of_faults_conf}, the share of configuration types across faults is similar to the share of configuration types across substations for both manufacturers. It should be noted that not all configurations listed in \ref{data} are present in those figures, since not all substations exhibit faults within the time range of sensor data. 

Furthermore, for each fault we marked a period prior to the incident report as a possible ‘anomaly’ onset if this was visible in the data and matched technically expected behaviour. The end of the anomaly was set to the timestamp of the subsequent maintenance action plus four hours, to ensure the action is completed and to allow the substation to stabilise with new settings or repaired/replaced components. These timestamps can be used for data preparation, such as filtering normal behaviour for the development of \acp{nbm} for early fault detection.

We also provide a set of pre-defined normal events. For these intervals, well-performing fault detection algorithms should not raise false alarms. The labelled fault and normal events can be combined into a balanced evaluation dataset for early fault detection. The normal events are selected across all seasons, primarily for substations in the reports list and several additional substations without faults. These intervals are free from maintenance tasks and faults.

In Table~\ref{tab:events_overview} an overview of the faults and normal events is provided.

\begin{table}[ht]
\centering
%\resizebox{\textwidth}{!}{%
\begin{tabular}{@{}llll@{}}
\toprule                                        & \textbf{\emph{M1}} & \textbf{\emph{M2}} & \textbf{Total} \\ \midrule
\# faults                                       & 33   & 40  & 73  \\ 
\# faults with a high monitoring potential      & 26   & 24  & 50  \\ 
\# faults with a low monitoring potential       & 5    & 11  & 16  \\ 
\# faults with unknown monitoring potential     & 2    & 5   & 7   \\ 
\# pre-defined normal events                    & 30   & 35  & 65  \\ 
\end{tabular}%
%}
\caption{Overview of the faults and normal events.  \emph{M1} and \emph{M2} refer to the two different control-unit manufacturers.}
\label{tab:events_overview}
\end{table}

\section{Methodology for the Evaluation of FDD methods}\label{method}
To illustrate how the PreDist dataset can be used to develop early fault detection models, we apply the EnergyFaultDetector, an open‑source Python package developed for automated anomaly detection in operational data from energy systems. The framework uses an \ac{ae} as \ac{nbm} to detect deviations from normal behaviour. It was originally developed by the authors for early fault detection in wind turbine data \cite{roelofs_autoencoder-based_2021} and has been adapted here for \acp{dhs}.

First, we select faults that are relevant for early fault detection and pre-process the data in Section~\ref{data_prep}. Next, the model used to detect anomalies is described in Section~\ref{ae_setup}. The evaluation metrics are explained in Section~\ref{evaluation} and the hyperparameter optimisation is discussed in Section~\ref{hyperparams}. Finally, we introduce the method used for root-cause analysis in Section~\ref{root-cause-analysis}

\subsection{Data selection and preparation}\label{data_prep}

Because we evaluate an early fault detection model, we select faults for which early detection may be feasible. We include only faults with at least 14 days of training data representing normal behaviour. We filter repeat reports (some faults have multiple reports) and consider only the first occurrence, as our goal is to detect the problem before the customer reports a fault. In addition, we remove reports concerning heat-meter battery replacement and reports where the cause of the fault is in the building heating system, i.e. not in the \ac{dhs}; the former are likely preventive, and the latter cannot be detected due to a lack of customer‑side measurements. The datasets leading to a fault are hereafter called `anomaly events', while datasets in which no anomalies are expected in the test data are used as is and are hereafter called `normal events’. For \emph{M1}, this leaves 29 reports and for \emph{M2} 26 reports. Of these reports, 22 and 15 have a high monitoring potential for \emph{M1} and \emph{M2}, respectively. All normal events (see Section~\ref{maintenance_data}) were selected for the evaluation: 35 for \emph{M1} and 30 for \emph{M2}.

For each selected anomaly and normal event, a training dataset is determined based on time series visualisations and fault descriptions. If a previous fault or maintenance measure changed normal behaviour, the training data starts after that point, otherwise, we select the last 2 years up to 2 weeks before the event to be predicted. If less data are available, we select data from the start of the time series until 2 weeks before the event. In some cases, the first days to weeks are removed because the commissioning of the substation involved frequent re-parametrisation of the control unit and no stable normal behaviour can be selected.

Next, the selected training datasets are prepared for training \ac{nbm} models. If any incident report or maintenance task occur in the training data, we filter out any possible anomalous behaviour as follows: If any anomalous behaviour is observed before an incident report, we filter out this period up to the subsequent maintenance action plus four hours to account for time it takes to complete the work. We exclude 48 h before reports without visible anomalies and assume the fault was fixed 24 h after the report if no maintenance is logged. %If no anomalies are observed before a report, we exclude 48 hours preceding the report from training to reduce the chance of including incipient faults. If no maintenance action was registered after a report, we assume the fault was fixed in the next 24 hours, as many of the incidents are addressed on the same day. 

Afterwards, features are selected for the models. Features that are constant throughout the training period and features that are missing in more than 80\% of the training period are dropped. Any other missing values are imputed with a mean value of the training data. Finally, features are scaled to zero mean and unit standard deviation.

\subsection{Autoencoder-based anomaly detection}\label{ae_setup}
The EnergyFaultDetector uses an \ac{ae} as \ac{nbm} at its core. The \ac{ae} is a type of neural network that learns to compress and reconstruct the input data.
%The \ac{ae} is a type of neural network that consists of two parts, the encoder and the decoder. The encoder maps the input $x$ to a compressed latent code $z$, and the decoder reconstructs $\tilde{x}$ from $z$. Trained on data representing normal behaviour, the \ac{ae} minimises the \ac{re} $RE_i=\tilde{x}_i-x_i$. 
As the \ac{ae} is trained on data representing normal behaviour, the model learns to reconstruct normal data and will show a larger \ac{re} for behaviour that deviates from normal behaviour. 

Anomalies are detected when the average \ac{re} (anomaly score) exceeds a threshold $t_{AE}$. %The anomaly score summarises the \acp{re} of the features. 
We consider two anomaly score types: the \ac{rmse} of $RE_i$ and the Mahalanobis distance,
\begin{equation}
\label{equ:Mahalanobis}
    \text{MD}_{x} =\sqrt{(\tilde{x} - x)^T {\Sigma}^{-1}(\tilde{x} - x)},
\end{equation}
where $\Sigma$ is the covariance matrix of the training errors. Unlike the \ac{rmse}, the Mahalanobis distance accounts for correlations between the \acp{re} of the individual features. However, since the covariance matrix $\Sigma$ can be hard to estimate reliably for high-dimensional data, we treat the score type as a hyperparameter. We set the threshold $t_{AE}$ to the 99th quantile of the training anomaly scores. 

Finally, to suppress point anomalies, we use a criticality counter $C$: it increases by one when an anomaly is detected during expected normal operation, decreases by one if no anomaly is detected, and remains constant otherwise (i.e., an anomaly detected during a maintenance task). An event is flagged as anomalous if the maximum criticality $C_{max}$ satisfies $C_{max} \ge C_{thr}$. Otherwise, the event is recognised as normal behaviour.

We train one EnergyFaultDetector model per event, using a generic model configuration per manufacturer.  We test three \ac{ae} variants: a default \ac{ae}, a conditional \ac{ae} with hour-of-day and day-of-week as conditional features, and a conditional “day-of-year” \ac{ae} that additionally uses day-of-year. Time-context features are encoded cyclically (sine/cosine) and concatenated to both the encoder input and the decoder input. %The day-of-year variant is evaluated separately, because a full year of training data is often unavailable, and it is not clear whether this feature helps the model in such cases or causes more false alarms.

\subsection{Evaluation}\label{evaluation}
We evaluate models using three metrics - accuracy, reliability and earliness - taking inspiration from the CARE score \cite{guck_care_2024}, but we do not compute the composite CARE value and we omit coverage because anomaly onsets are generally unknown in this dataset. A limitation of CARE is its dependence on ground truth derived from annotated onsets. This only affects coverage and earliness, both of which rely on the anomaly onset time.

\begin{comment}
\begin{figure}
    \centering
    \includegraphics[width=0.8\linewidth]{img/anomaly_onset_to_report_timeline.jpg}
    \caption{Timeline illustrating anomaly onset, fault occurrence, and customer incident report. The intervals $t_F$ (onset $\rightarrow$ fault), $t_R$ (fault $\rightarrow$ report), and $t_A = t_F + t_R$ (onset $\rightarrow$ report) frame the actionable window for early detection.}
    \label{fig:anomaly_onset_timeline}
\end{figure}
Figure~\ref{fig:anomaly_onset_timeline} illustrates the operational timeline: $t_F$ is the time from anomaly onset to fault manifestation (e.g., pump failure), $t_R$ is the time from fault to incident report (e.g., no heat), and $t_A = t_F + t_R$ is the time from anomaly onset to report. $t_F$ depends on the failure mode, ranging from weeks (e.g., heat‑exchanger fouling) to effectively zero (e.g., external leakage), while $t_R$ depends on the thermal inertia of the system and customer behaviour. The ground-truth for anomaly onset and fault manifestation is unknown. However, from an O\&M perspective, we are ultimately interested in $t_A$, i.e. detecting the earliest actionable anomaly before the customer complains. 
\end{comment}

Without a ground‑truth onset coverage cannot be computed, and earliness requires a redefinition. From an O\&M perspective, we are ultimately interested in detecting the earliest actionable anomaly before the customer complains. Therefore, we remove dependency on annotated onsets by defining the detection time $t_{\text{detect}}$ as the earliest timestamp in the test window at which the criticality crosses $C_{thr}$. We normalise earliness by a window length $W$, the period in which detection remains actionable:
\begin{equation}
    E = \max\left(0, \min\left(1, \frac{t_{\text{report}} - t_{\text{detect}}}{W}\right)\right),
\end{equation}
where $t_{\text{report}}$ is the incident report timestamp. If no detection occurs within the test data (i.e., criticality never crosses $C_{thr}$), we set $E=0$ for that event. The desired detection time $W$ should be set to a value which makes detections actionable and is use-case specific. In this case we set $W = 24\,\mathrm{h}$.

\begin{figure}
    \centering
    \includegraphics[width=\linewidth]{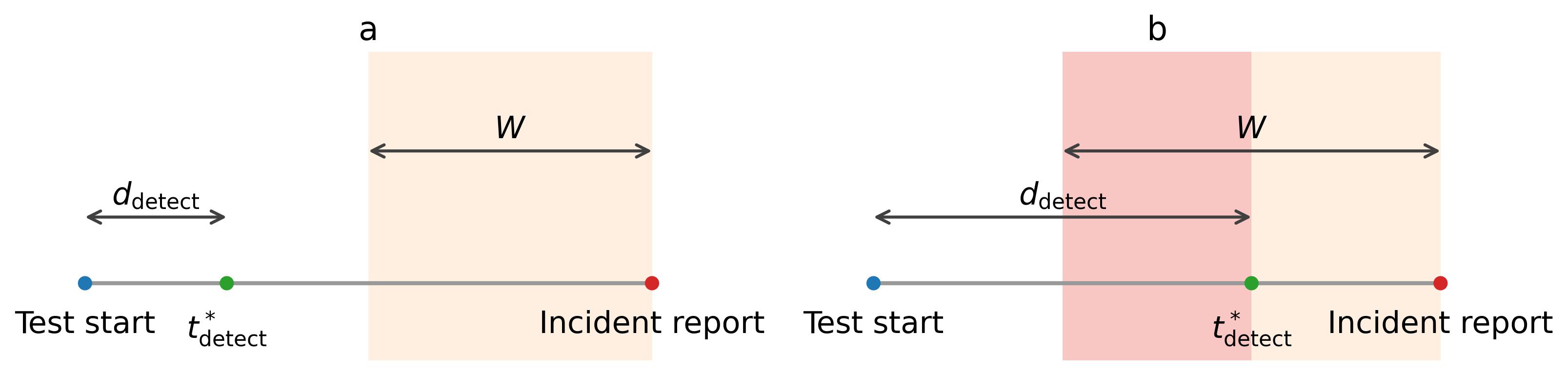}
    \caption{Schematic timeline, illustrating detection delay $d_{\text{detect}}$, earliest detection time $t^*_{\text{detect}}$ and desired lead time $W$ before an incident report. In Figure b $d_{\text{detect}}$ and $W$ overlap, so earliness can never reach 1.}
    \label{fig:detection_delay}
\end{figure}

Note that $W$ should not be longer than the shortest anomaly event (if an anomaly onset is known) or the test window length, so the event can be detected within the test window. It should also ideally not exceed the maximum achievable lead time $L^* = t^*_{\text{detect}} - t_{\text{report}}$, where $t^*_{\text{detect}}$ is the earliest detection time. For a given $C_{thr}$, the detection delay is $d_{\text{detect}} = C_{thr}/f$, where $f$ is the sampling frequency of the data. The earliest detection time is then $t^*_{\text{detect}} = t_{\text{test\_start}} + d_{\text{detect}}$. This ensures it is possible to reach $E=1$ and all events are compared against the same maximum score. Figure~\ref{fig:detection_delay} shows a schematic timeline visualising $W$ and $d_{\text{detect}}$.% where Figure~\ref{fig:detection_delay}a shows suitable values for $W$ and $C_{thr}$, while Figure~\ref{fig:detection_delay}b shows overlapping $d_{\text{detect}}$ and $W$, so a fault detection model can never get $E = 1$.

Reliability $R$ is kept as an eventwise F$_\beta$ score, where we set $\beta=0.5$, to favour precision over recall. We compute eventwise $F_{0.5}$ by counting a true positive if an anomalous event’s criticality crosses $C_{thr}$ within its test window, a false negative otherwise, and a false positive if a normal event crosses $C_{thr}$ (true negative otherwise). We use $R$ as our main metric to set the $C_{thr}$ and evaluate the models. In addition, the $\overline{E}$ is reported, to show how long before $t_{\text{report}}$ a fault can be detected.

To evaluate the models' ability to recognise normal behaviour correctly we evaluate the pointwise accuracy A as in the CARE score and report the average over all normal events.

% Test data
We evaluate each event on a fixed test window of 7 days. For anomalous events, this window ends at the event report time. One exception to this rule is a fault that is visible for more than 7 days. Here, we evaluate only the first 7 days, since an effective early detection system should detect the fault within that period.
An anomalous event is considered `detected before the customer report' if the criticality threshold $C_{thr}$ is crossed at least once within the 7‑day test window preceding the incident report timestamp.

\subsection{Hyperparameter optimisation}\label{hyperparams}
We tune hyperparameters of the \ac{ae} models in two phases. In the first phase we tune the \ac{ae}’s architecture to learn different types of normal substation behaviour well. %For this, we tuned the number of hidden layers, units per layer, the learning rate and the activation functions of the hidden and output layers. 
We minimise validation \ac{mse} using a 80/20 train–validation split within each event. %We train one model per event and optimise the average validation \ac{mse} across events. 

In the second phase, we calibrate the fault detection performance. We tune hyperparameters that affect detection performance using the earliness score on anomalous events with a provisional $C_{thr} = 36$, while keeping pointwise accuracy on normal events $\ge 0.97$ to limit false alarms. %We tune the input noise (denoising regularisation), the anomaly score type (\ac{rmse} vs. Mahalanobis), and latent dimension. The latent dimension is tuned again, since the \ac{ae} often needs a large latent dimension to reconstruct as precise as possible (needed to minimise the \ac{mse}), but a smaller one helps to be more sensitive to deviating behaviour.
%The noise parameter functions as a regularisation hyperparameter. It adds Gaussian noise to the input data to make the \ac{ae} less sensitive to small changes in the time series, effectively changing the \ac{ae} to a denoising \ac{ae}.

Table~\ref{tab:hyperparams} reports the final hyperparameters for each manufacturer. Because each substation has a different number of input features, we use a relative value of the latent dimension, reported as a fraction of the input features and then rounded to an integer.

\begin{table}[ht]
\centering
%\resizebox{.8\textwidth}{!}{%
\begin{tabular}{@{}lll@{}}
\toprule                 & \textbf{\emph{M1}}      & \textbf{\emph{M2}}      \\ \midrule
\# units (hidden layers) & 64, 32         & 64, 32         \\
latent space dimension   & 0.65           & 0.25           \\
learning rate            & 0.00045        & 0.00053        \\
noise                    & 0.05           & 0.15           \\
batch size               & 256            & 256            \\
anomaly score            & Mahalanobis    & RMSE           
\end{tabular}%
%}
\caption{Hyperparameters of the fault detection models for each manufacturer. Noise is used as a regularisation hyperparameter. It represents Gaussian noise added to the input data of the \ac{ae}. \emph{M1} and \emph{M2} refer to the two different control-unit manufacturers.}
\label{tab:hyperparams}
\end{table}

We perform hyperparameter optimisation on the default \ac{ae} setup, then reuse the same configuration for the conditional \acp{ae}. The size of the latent dimension applies only to actual input features (measurements and setpoints). Conditional features are not compressed. Because several fault types occur only once, we do not hold out a separate test set for hyperparameter optimisation, which may lead to overfitting. To mitigate this, we keep a single configuration per manufacturer across all substations. Per‑substation tuning would likely improve scores but increases overfitting risk.

For the criticality threshold $C_{\mathrm{thr}}$, we performed a grid search over integer values from 1 to 100 using 5-fold stratified cross-validation by event label, selecting the value that maximised reliability averaged across folds. The selected $C_{\mathrm{thr}}$ per model and manufacturer is reported in Table~\ref{tab:thresholds}. 
\begin{comment}
As an example, the threshold selection for the default \ac{ae} on the \emph{M1} dataset, is shown in Figure~\ref{fig:threshold_selection_default_m1}.
\begin{figure}
    \centering
    \includegraphics[width=1\linewidth]{img/M1_threshold_selection_default_ae.jpg}
    \caption{Threshold selection of the default \ac{ae} model for the \emph{M1} dataset using a stratified k-fold with $k=5$.}
    \label{fig:threshold_selection_default_m1}
\end{figure}
\end{comment}

\begin{table}[ht]
\centering
\begin{tabular}{lcc} 
\toprule Model             & \emph{M1} & \emph{M2} \\
\midrule 
Default \ac{ae}            & 17        & 24        \\
Conditional \ac{ae}        & 12        & 19        \\
Day-of-year \ac{ae}        & 9         & 8         \\
\end{tabular}%
\caption{Criticality threshold for all three model variants tested. \emph{M1} and \emph{M2} refer to the two different control-unit manufacturers.}
\label{tab:thresholds}
\end{table}

\subsection{Root‑cause analysis}\label{root-cause-analysis}
For post‑hoc root‑cause analysis on detected anomalies we apply ARCANA, which is integrated in the EnergyFaultDetector. ARCANA is a feature attribution method developed for \acp{ae} that helps to find a possible root cause for detected anomalies \cite{roelofs_autoencoder-based_2021}. The method is based on the assumption that faults cause a deviation in only a small subset of features. For each input vector $x$, ARCANA finds a sparse bias vector $x_{\mathrm{bias}}$ such that the corrected input $x_{\mathrm{corr}}=x+x_{\mathrm{bias}}$ is reconstructed with a low \ac{mse} while keeping $x_{\mathrm{bias}}$ sparse. To find $x_{\mathrm{bias}}$, ARCANA minimises 
\begin{equation}
    \mathcal{L}=(1-\alpha)\,\tfrac{1}{2}\|x_{\mathrm{corr}}-\text{AE}(x_{\mathrm{corr}})\|_{2}^{2}
    +\alpha\|x_{\mathrm{corr}}-x\|_{1},
\end{equation}
where $\text{AE}$ is the \ac{ae} model. In our experiments we use \(\alpha=0.8\) and initialise \(x_{\mathrm{bias}}\) with the feature-wise reconstruction error $RE$. 

We compute feature importances by aggregating the bias vectors over a detected anomaly window: first we average the absolute bias values over the window to obtain unnormalised importances, then we normalise them so that their sum equals one. Features are ranked by these normalised importances and the top‑3 features are reported as candidate root causes. These features are visualised together with their original and reconstructed time series to facilitate operator interpretation.

\section{Results and discussion}\label{results}
As mentioned in Section~\ref{ae_setup}, we compared three model variants: a default \ac{ae} model, a conditional \ac{ae} that uses day-of-week and hour-of-day features as conditional features, and a conditional \ac{ae} which also takes the day-of-year into account, which we refer to as the `day-of-year \ac{ae}'. 

\subsection{Overall results}
\begin{figure}
    \centering
    \includegraphics[width=1\linewidth]{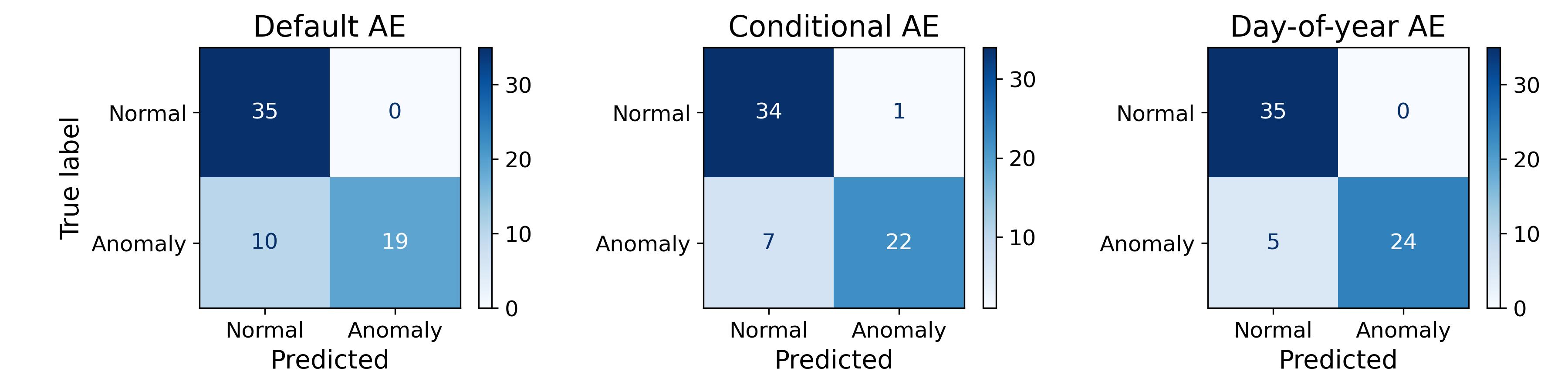}
    \caption{Results for the three models for the \emph{M1} dataset.}
    \label{fig:confusion_matrix_m1}
\end{figure}

\begin{figure}
    \centering
    \includegraphics[width=1\linewidth]{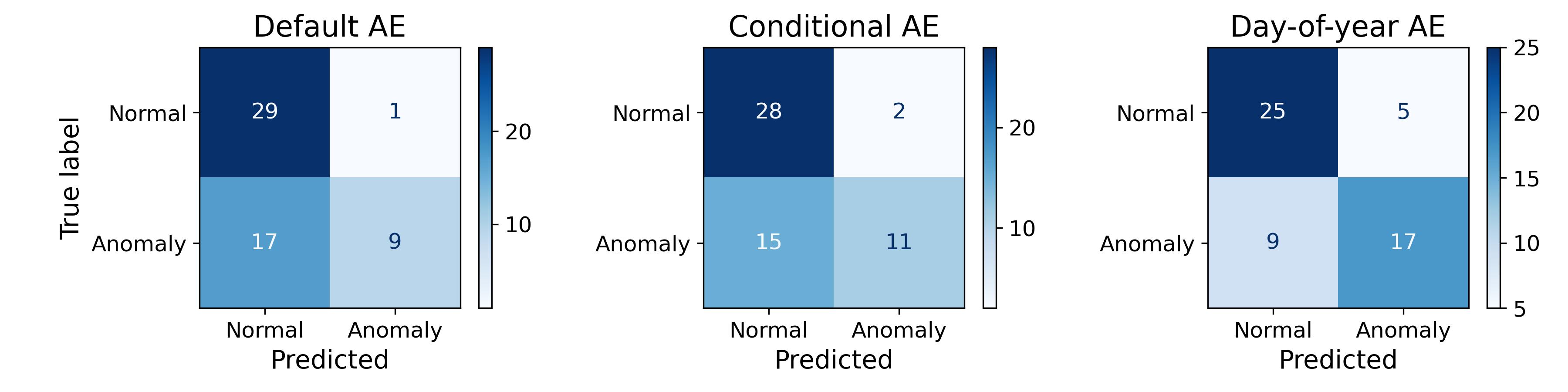}
    \caption{Results for the three models for the \emph{M2} dataset.}
    \label{fig:confusion_matrix_m2}
\end{figure}

The confusion matrices for all six models are shown in Figures~\ref{fig:confusion_matrix_m1} and \ref{fig:confusion_matrix_m2}. For both \emph{M1} and \emph{M2} datasets the conditional \ac{ae} models perform slightly better than default \ac{ae} models. The day-of-year \ac{ae} seems to perform best for both datasets, however, this is probably caused by the small amount of training data available for anomalous events. In the case of \emph{M2}, an average of 310 days of training data is available for anomalous events, whereas for normal events an average of 704 days is available. For \emph{M1}, the averages are 588 for anomalies and 576 days for normal events.
Although accounting for day-of-year can help, short training periods mean that detections may reflect seasonal changes rather than faults.

An overview of accuracy on normal events and reliability is shown in Tables~\ref{tab:results_m1} and \ref{tab:results_m2} for datasets \emph{M1} and \emph{M2}, while Table~\ref{tab:results_overall} shows the results averaged over both datasets. Note that the reliability score is overestimated, since we use a balanced dataset to calculate the scores. In reality, the fraction of faults is much lower, which might lead to an overestimation of the precision and thereby the reliability reported. Because there is no canonical normal‑to‑fault ratio, we retain a 50/50 split. To aid interpretation, we also report eventwise precision and recall and note that precision is likely lower under realistic class imbalance.
\begin{table}[ht]
\centering
\resizebox{\textwidth}{!}{%
\begin{tabular}{lcccccc} 
\toprule Model             & A               & R    & Precision & Recall  & E                 & L (d) \\
\midrule 
Default \ac{ae}            & 0.98 $\pm$ 0.01 & 0.90 & 1.00      & 0.66    & 0.59 $\pm$ 0.16   & 2.8 $\pm$ 1.2    \\  % 68h
Conditional \ac{ae}        & 0.98 $\pm$ 0.01 & 0.88 & 0.95      & 0.69    & 0.73 $\pm$ 0.17   & 4.0 $\pm$ 1.0   \\  % 96h
Day-of-year \ac{ae}        & 0.98 $\pm$ 0.01 & 0.94 & 1.00      & 0.76    & 0.81 $\pm$ 0.16   & 4.9 $\pm$ 0.9   \\  % 118h
\bottomrule
\end{tabular}
}
\caption{Results for the three models on the \emph{M1} dataset. Reliability is eventwise $F_{0.5}$, precision and recall are also computed eventwise. Accuracy is the average pointwise accuracy on normal events. Earliness is the average earliness score on anomaly events. The lead time $L$ is an average over detected faults only. For accuracy, earliness, and lead time, 95\% confidence intervals across events are reported.}
\label{tab:results_m1} 
\end{table}
\begin{table}[ht]
\centering
\resizebox{\textwidth}{!}{%
\begin{tabular}{lcccccc}
\toprule Model             & A               & R    & Precision & Recall  & E                 & L (d) \\
\midrule 
Default \ac{ae}            & 0.98 $\pm$ 0.01 & 0.68 & 0.90      & 0.35    & 0.25 $\pm$ 0.16   & 2.7 $\pm$ 2.1    \\  % 65h
Conditional \ac{ae}        & 0.98 $\pm$ 0.01 & 0.71 & 0.85      & 0.42    & 0.35 $\pm$ 0.19   & 3.7 $\pm$ 1.8    \\  % 88h
Day-of-year \ac{ae}        & 0.96 $\pm$ 0.01 & 0.75 & 0.77      & 0.65    & 0.64 $\pm$ 0.19   & 5.4 $\pm$ 0.9    \\  % 130h
\bottomrule 
\end{tabular} 
}
\caption{Results for the three models on the \emph{M2} dataset. Reliability is eventwise $F_{0.5}$, precision and recall are also computed eventwise. Accuracy is the average pointwise accuracy on normal events. Earliness is the average earliness score on anomaly events. The lead time $L$ in days is an average over detected faults only.  For accuracy, earliness, and lead time, 95\% confidence intervals across events are reported.} 
\label{tab:results_m2}
\end{table}
\begin{table}[ht]
\centering
\resizebox{\textwidth}{!}{%
\begin{tabular}{lcccccc}
\toprule Model             & A               & R    & Precision & Recall  & E                & L (d) \\
\midrule 
Default \ac{ae}            & 0.98 $\pm$ 0.01 & 0.82 & 0.97      & 0.51    & 0.39 $\pm$ 0.13  & 3.2 $\pm$ 1.0    \\  % 65h
Conditional \ac{ae}        & 0.98 $\pm$ 0.01 & 0.83 & 0.92      & 0.60    & 0.54 $\pm$ 0.13  & 4.0 $\pm$ 1.0    \\  % 88h
Day-of-year \ac{ae}        & 0.98 $\pm$ 0.01 & 0.86 & 0.89      & 0.75    & 0.72 $\pm$ 0.12  & 5.4 $\pm$ 0.8    \\  % 130h
\bottomrule 
\end{tabular} 
}
\caption{Results for the three models on both datasets (micro-averages). Reliability is eventwise $F_{0.5}$, precision and recall are also computed eventwise. Accuracy is the average pointwise accuracy on normal events. Earliness is the average earliness score on anomaly events. The lead time $L$ in days is an average over detected faults only. For accuracy, earliness, and lead time, 95\% confidence intervals across events are reported.} 
\label{tab:results_overall}
\end{table}
In general, the conditional \ac{ae}-based models outperform the default \ac{ae}  on reliability and earliness at comparable accuracy. 
The earliness score and the average lead time are better for the conditional \ac{ae} models, due to the optimal threshold being lower, i.e. the models are more sensitive to a change in behaviour than the default \ac{ae} (see Table~\ref{tab:thresholds}). We also find that the day-of-year \ac{ae} model detects more faults earlier, but with a much lower precision. This is most likely due to the earlier mentioned short training periods: this model variant often cannot reconstruct the outside temperature, and the accuracy for normal events is slightly lower for the \emph{M2} dataset. We therefore regard the conditional \ac{ae} as the best overall model.

\subsection{High-monitoring potential}
Next, we compare the model performance on earliness and reliability across all faults with the model performance for faults with a high monitoring potential, as defined in Section~\ref{maintenance_data}, only. Some faults rated a high monitoring potential are present from commissioning onwards and are therefore not detectable as a deviation from normal behaviour. These are, for example, `incorrect parametrisation of the control unit' and `misplacement of the outdoor temperature sensor'. These are excluded from the high monitoring potential faults, as these should be addressed via installation checks or auto-commissioning \cite{guevara_bastidas_prioritisation_2025}. This leaves 14 faults for which early fault detection is possible with a high monitoring potential for both datasets.

\begin{figure}
    \centering
    \includegraphics[width=\linewidth]{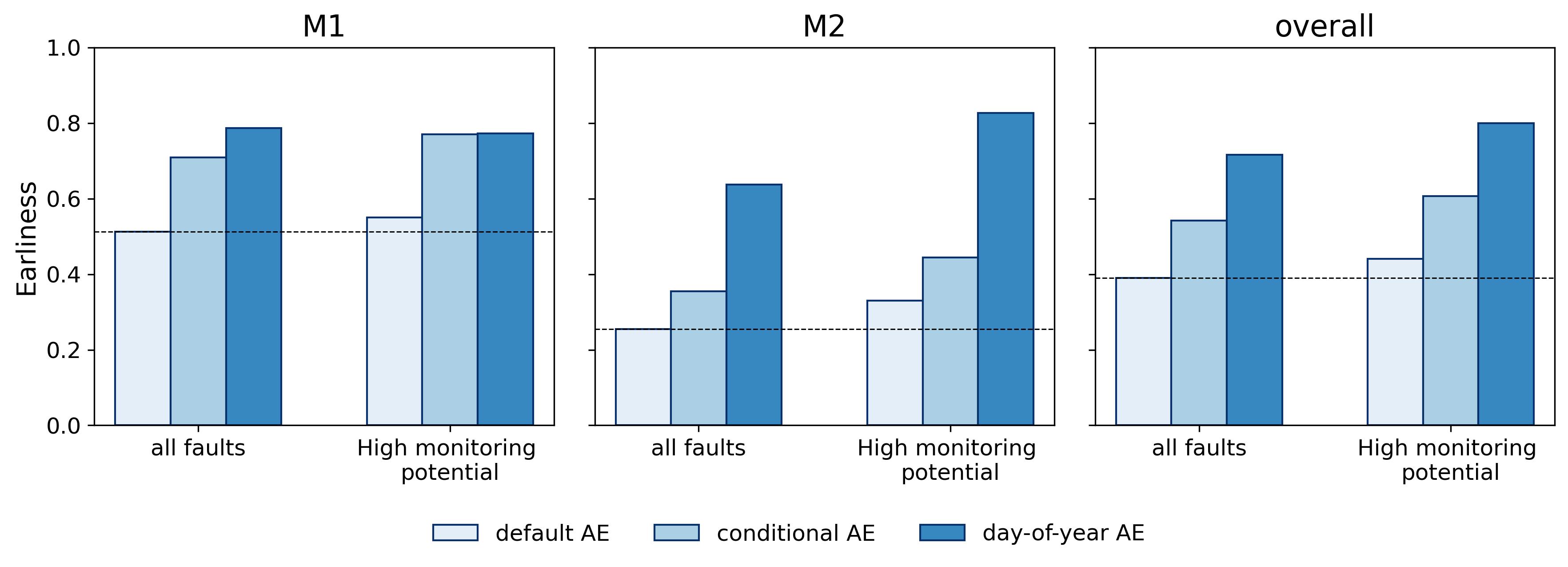}
    \caption{Comparison of the earliness score of the three models for both datasets between all faults and the faults with high monitoring potential. The reference dotted line is placed at the earliness of default \ac{ae} on all faults.}
    \label{fig:model_comparison}
\end{figure}

\begin{figure}
    \centering
    \includegraphics[width=\linewidth]{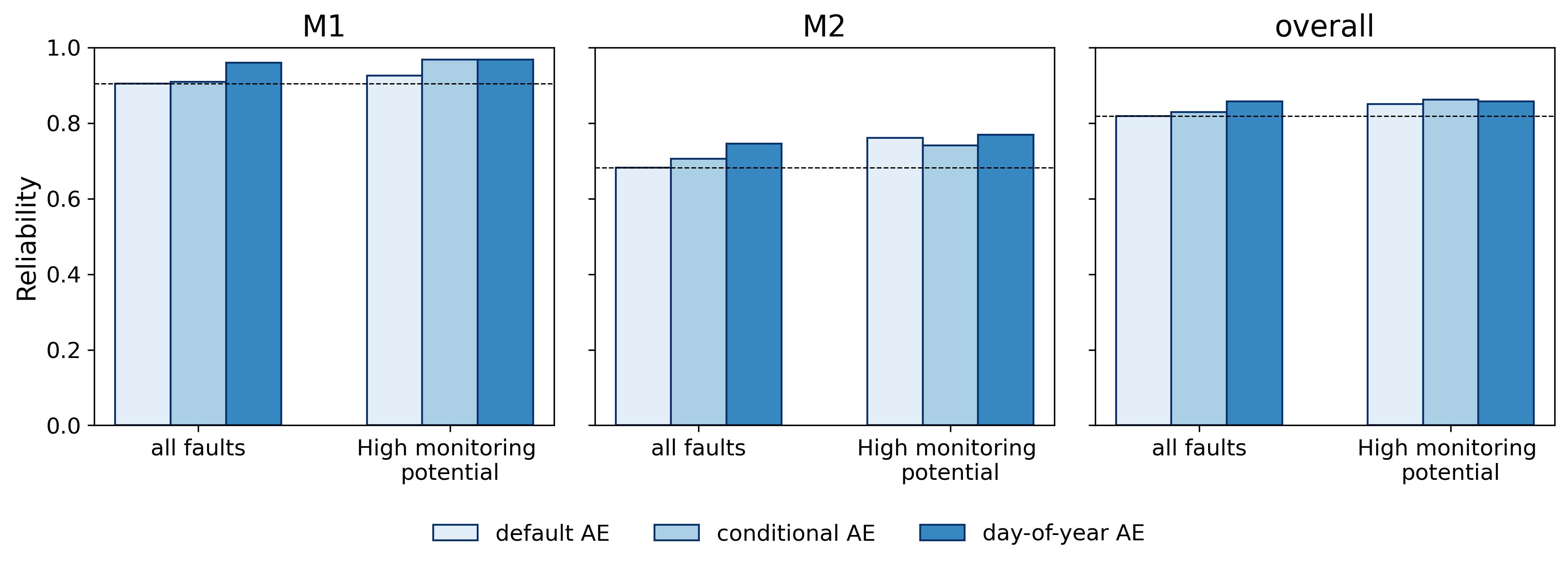}
    \caption{Comparison of the reliability score (eventwise $F_{0.5}$) of the three models for both datasets between all faults and the faults with high monitoring potential. The reference dotted line is placed at the reliability of default \ac{ae} on all faults}
    \label{fig:model_comparison_reliability}
\end{figure}

The results are shown in Figures~\ref{fig:model_comparison} and \ref{fig:model_comparison_reliability} for both datasets.
The faults with a high monitoring potential should be detectable even before the actual fault and we therefore expect the models to be able to detect these faults earlier. Also, since a high monitoring potential indicates that the faults should generally be detectable without additional sensors, a higher reliability is expected. 

\subsection{Use cases}
Using three example use-cases, we demonstrate how the criticality and ARCANA feature importances can be used to support early fault detection and diagnosis. For each case, we analyse the seven days preceding the incident report, compare the three models, and visualise the top‑3-ranked ARCANA features alongside their reconstructions to guide interpretation.

The input signals used for the use cases are listed in Table~\ref{tab:usecase_inputs}. Available features differ across \ac{dhs} due to different configurations (one or two heat circuits, space heating or \ac{dhw} or both, etc.). The complete list of available features across all substations in the PreDist dataset are provided with the dataset on Zenodo.

\begin{table}[h!]
\centering
\resizebox{\textwidth}{!}{%
\begin{tabular}{l p{0.5\textwidth} l ccc}
\toprule
Side      & Feature                             & Unit & Use Case 1 & Use Case 2 & Use Case 3 \\
\midrule
---       & outdoor temperature                 & °C   & \checkmark & \checkmark & \checkmark \\
\midrule
primary   & network supply temperature          & °C   & \checkmark & \checkmark & \checkmark \\
primary   & network return temperature          & °C   & \checkmark & \checkmark & \checkmark \\
primary   & flow (energy meter)                 & m3/h & \checkmark & \checkmark & \checkmark \\
primary   & heat power (energy meter)           & kW   & \checkmark & \checkmark & \checkmark \\
primary   & HC1 return temperature              & °C   & \checkmark & \checkmark & \checkmark \\
primary   & HC1 return temperature setpoint     & °C   & ---        & ---        & \checkmark \\
primary   & HC1 control valve position setpoint & \%   & ---        & ---        & \checkmark \\
\midrule
secondary & HC1 supply temperature              & °C   & \checkmark & \checkmark & \checkmark \\
secondary & HC1 supply temperature setpoint     & °C   & \checkmark & \checkmark & \checkmark \\
\midrule
secondary & HC2 supply temperature              & °C   & \checkmark & ---        & --- \\
secondary & HC2 supply temperature setpoint     & °C   & \checkmark & ---        & --- \\
\midrule
secondary & upper storage temperature           & °C   & \checkmark & ---        & \checkmark \\
secondary & lower storage temperature           & °C   & \checkmark & ---        & \checkmark \\
\bottomrule
\end{tabular}%
}
\caption{Descriptions of input features used across the three example use cases. Check marks indicate which features were available in each case. Available features differ per substation configuration. HC1 and HC2 refer to heat circuits 1 and 2. Storage temperatures refer to a either \ac{dhw} tank (Use Case 1) or a heating buffer (Use Case 3).}
\label{tab:usecase_inputs}
\end{table}

% Event 49, substaion 18, 	configuration_type:	combined space heating and DHW
\subsubsection{Example 1 - \emph{M1} - no \ac{dhw}}
\begin{figure}
    \centering
    \includegraphics[width=\linewidth]{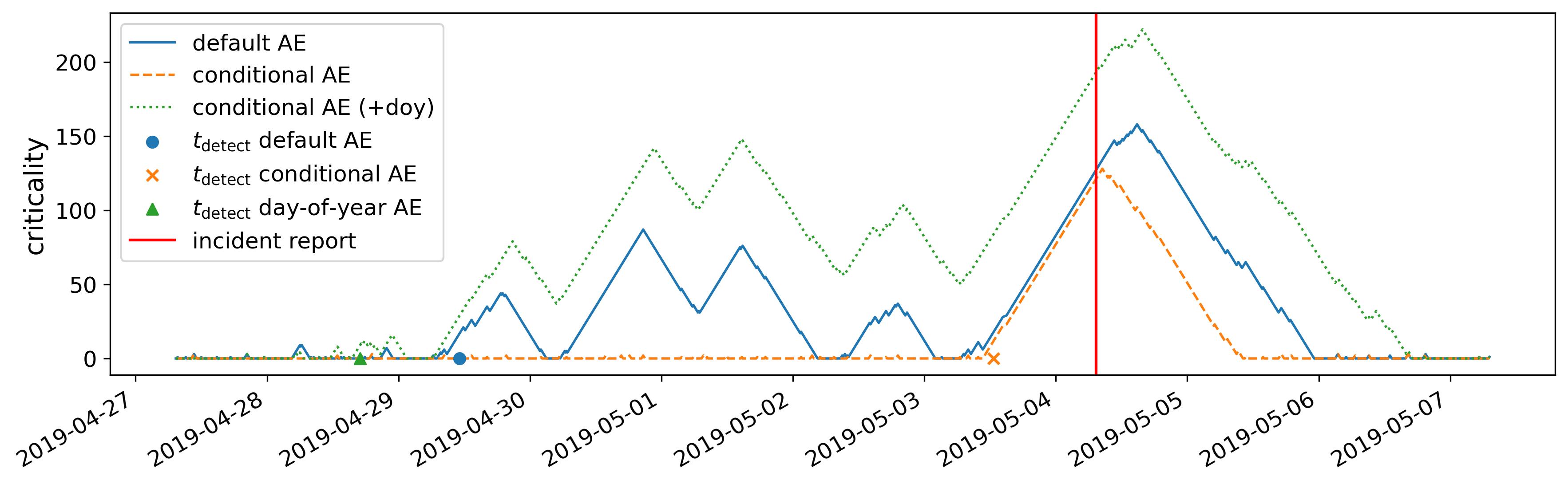}
    \caption{Criticality of the three model variants in the 7 days preceding an incident report of dataset \emph{M1} for `no \ac{dhw}' (example use-case 1). Multiple deviations from learned behaviour are detected before the report. All three models detect a problem 24 hours before the report is made, with the default and conditional \ac{ae} models detecting a change in behaviour 5-6 days before the report. After the problem is fixed, the criticality reduces to zero again.}
    \label{fig:example_1_criticality}
\end{figure}

\begin{figure}
    \centering
    \includegraphics[width=\linewidth]{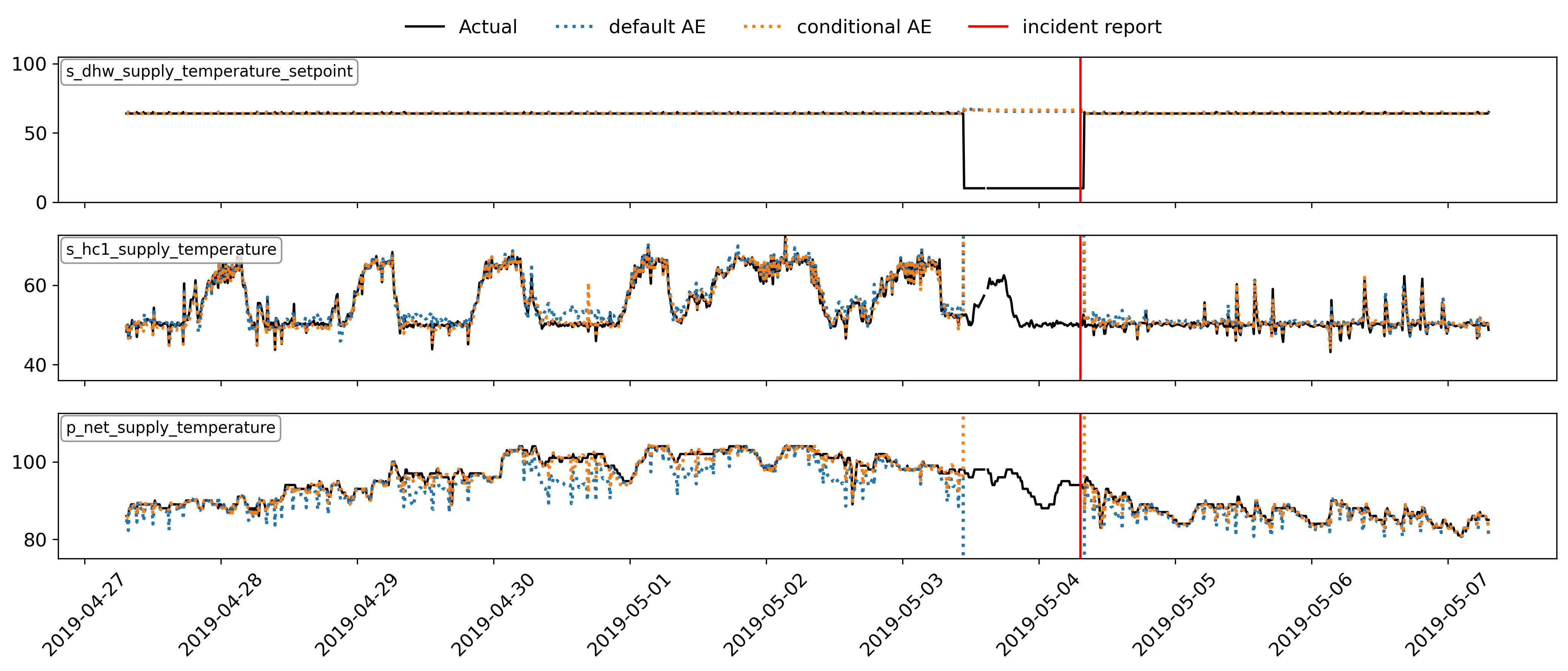}
    \caption{Top-3 deviating features for example use-case 1 according to the conditional \ac{ae}. The reconstructions of the default \ac{ae} and conditional \ac{ae} alongside the actual values are shown. Features are sorted by ARCANA importance in descending order. The vertical red line represents the time of the incident report.}
    \label{fig:example_1_topk}
\end{figure}

\begin{figure}
    \centering
    \includegraphics[width=\linewidth]{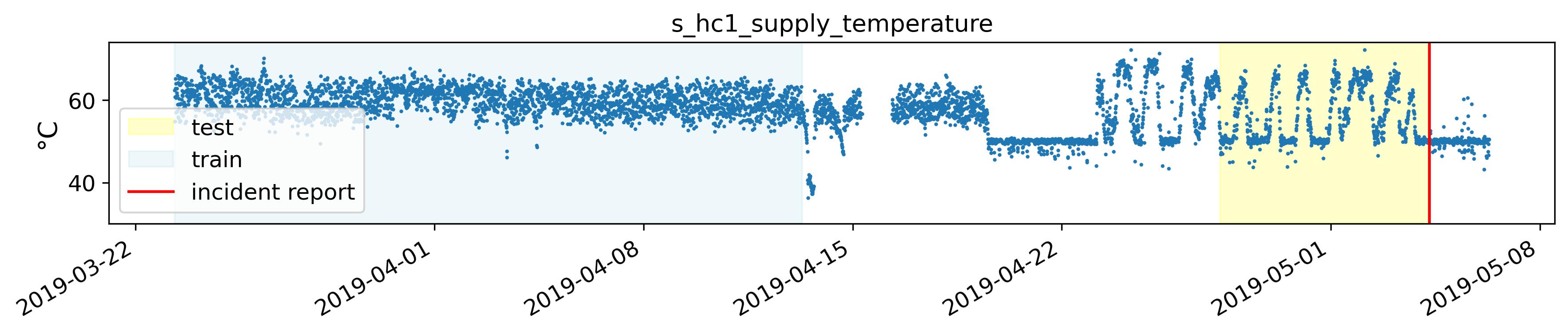}
    \caption{Secondary supply temperature of the \ac{dhs} in use-case 1 during training and testing. The behaviour of the secondary supply temperature changes after the training period.}
    \label{fig:example_1_hc1_supply}
\end{figure}

In this use-case, the substation has a combined space heating and DHW configuration (\texttt{SH + DHW}, see Section~\ref{data}). The customer called because of lack of hot water. It turned out to be an operating error where the \ac{dhw} controller was set to night mode, leading to a very low setpoint for the \ac{dhw} storage temperatures. Figure~\ref{fig:example_1_criticality} shows a clear increase in the criticality trends for all three models approximately 24 hours before the report. For all three models, ARCANA assigns the highest importance to the secondary supply temperature setpoint for the \ac{dhw} circuit, which drops abruptly to 10 \degree C and returns to 65 \degree C after the controller is reset to automatic, as seen in Figure~\ref{fig:example_1_topk}. The models fail to reconstruct this step change, which causes a high anomaly score. The default \ac{ae} and the day-of-year \ac{ae} also detect anomalies in the days before the controller setting was changed. The second highest importance is assigned to the secondary supply temperature of heat circuit 1 for the default and conditional \acp{ae}. This turns out to be the cause for the default \ac{ae} and the day-of-year \ac{ae} to detect anomalies before the controller setting was changed. This is due to a changed behaviour in the secondary supply temperature in heat circuit 1 as seen in Figure~\ref{fig:example_1_hc1_supply}, from more or less constant to daily oscillations. 


% Event 64, substation 6, 	configuration_type:	combined space heating and DHW
\subsubsection{Example 2 - \emph{M1} - insufficient heat}
\begin{figure}
    \centering
    \includegraphics[width=\linewidth]{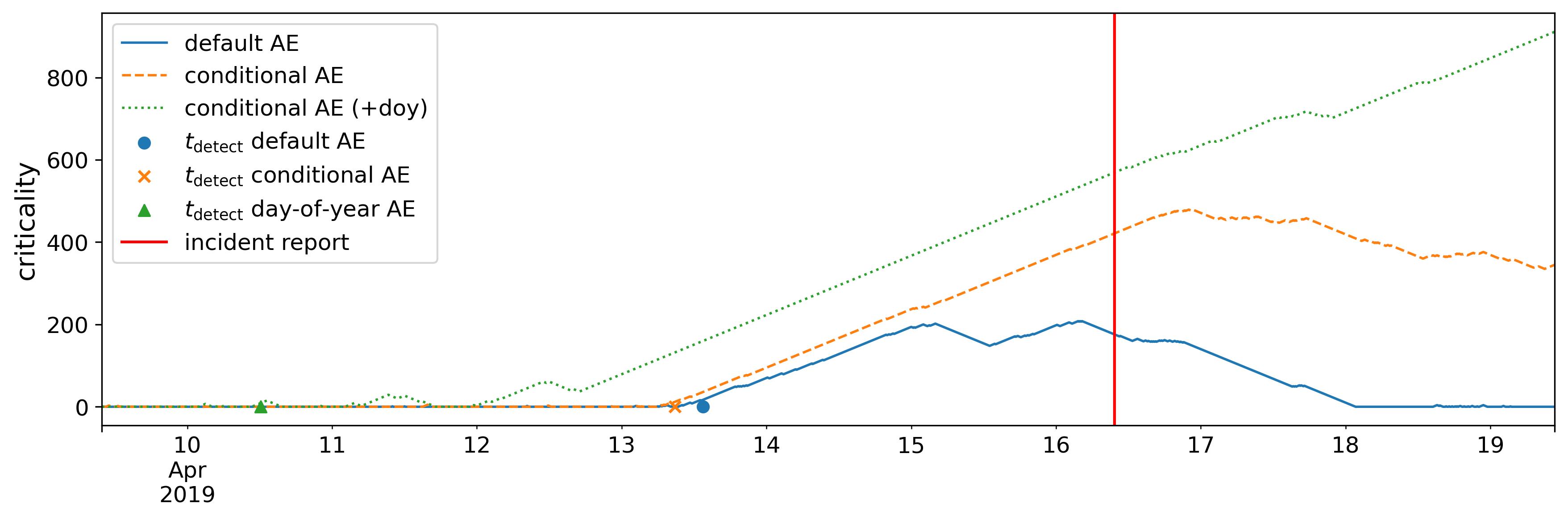}
    \caption{Criticality of the three model variants in the 7 days preceding an incident report of dataset \emph{M1} (example use-case 2). The criticality rises 3 to 4 days before the report is made for across all models. For the default and conditional \ac{ae} models, the criticality decreases after the fault is addressed.}
    \label{fig:example_2_criticality}
\end{figure}

\begin{figure}
    \centering
    \includegraphics[width=\linewidth]{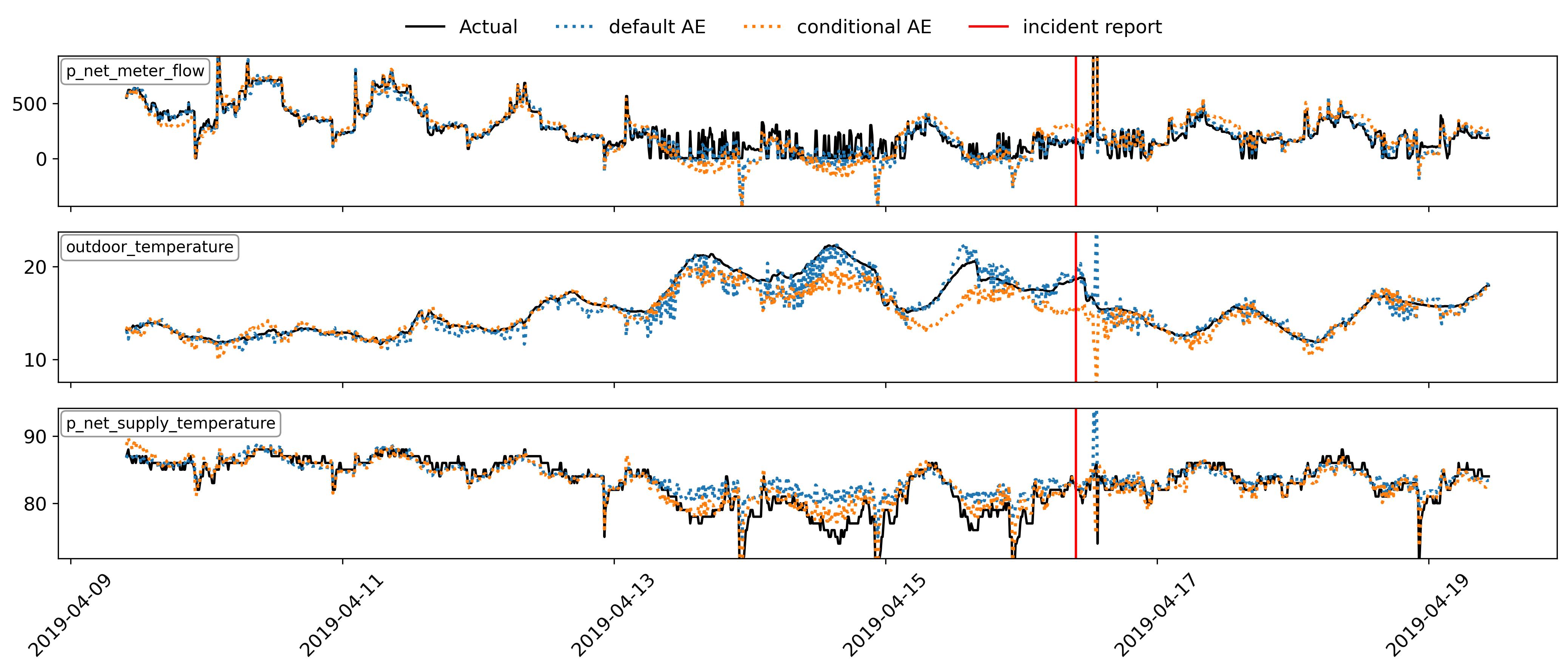}
    \caption{Top-3 deviating features for example use-case 2 according to the conditional \ac{ae}. The reconstructions of the default \ac{ae} and conditional \ac{ae} alongside the actual values are shown. Features are sorted by ARCANA importance in descending order. The vertical red line represents the time of the incident report.}
    \label{fig:example_2_topk}
\end{figure}

For the second example the problem was a defective differential pressure regulator. The substation has a combined space heating and DHW configuration (\texttt{SH + DHW}, see Section~\ref{data}). The criticality trends, shown in Figure~\ref{fig:example_2_criticality}, rise 3–4 days before the report for all model variants. While the criticality of the default and conditional \ac{ae} models decreases after the problem is addressed, the criticality according to the day-of-year \ac{ae} keeps increasing. As the training period was short, and the problem occurs in April, the day-of-year \ac{ae} model likely detects a seasonal change as well, instead of an actual problem. 
ARCANA identifies the primary‑side flow as the main driver for the detected anomalies. In Figure~\ref{fig:example_2_topk} the time series of the ARCANA top-3 features are shown, alongside the reconstruction of the default \ac{ae} and the conditional \ac{ae}. It can be seen that the flow is near-zero, which the models cannot reconstruct properly. The intermittent zero-flow values are a characteristic of a failing differential pressure regulator that intermittently restricts flow.


% Event 20, substation 50 (M2), 	configuration_type:	only space heating including buffer tank
\subsubsection{Example 3 - \emph{M2} - no heat}
\begin{figure}
    \centering
    \includegraphics[width=\linewidth]{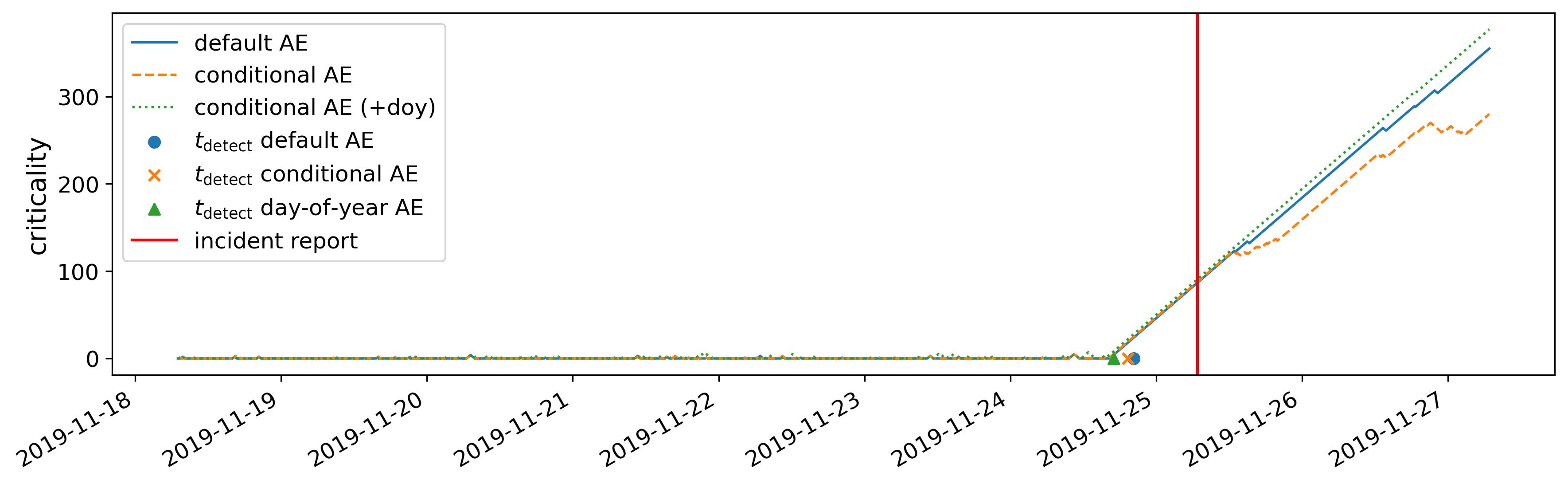}
    \caption{Criticality of the three model variants in the 7 days preceding an incident report for `no heat' of dataset \emph{M2} (example use-case 3).  All three models detect anomalies 10 hours before the report is made.}
    \label{fig:example_3_criticality}
\end{figure}

\begin{figure}
    \centering
    \includegraphics[width=\linewidth]{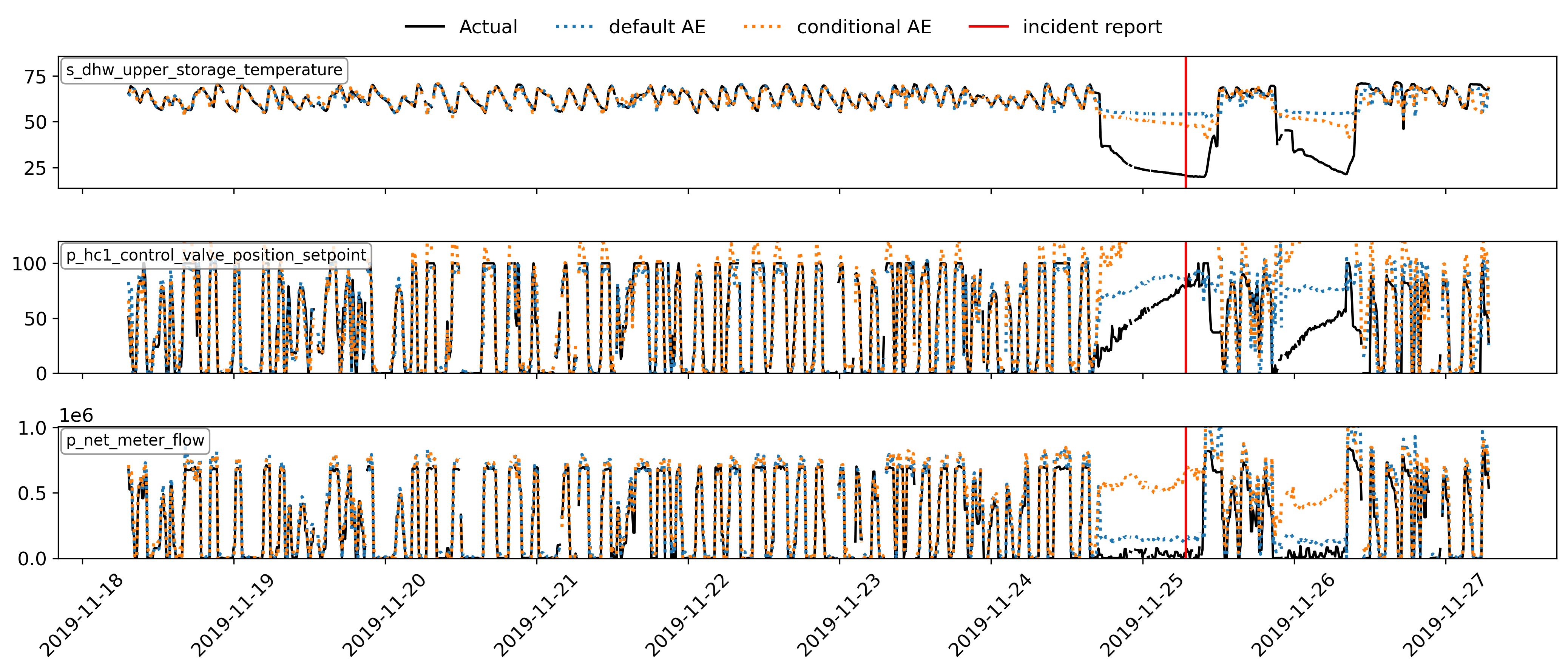}
    \caption{Top-3 deviating features for example use-case 3 according to the conditional \ac{ae}. The reconstructions of the default \ac{ae} and conditional \ac{ae} alongside the actual values are shown. Features are sorted by ARCANA importance in descending order. The vertical red line represents the time of the incident report.}
    \label{fig:example_3_topk}
\end{figure}

The subject of this use-case is a substation with only one heating circuit including a buffer storage tank (\texttt{SH with buffer tank}, see Section~\ref{data}). In this event the charging of the storage tank became very slow, leading to no space heating. The cause was a malfunctioning charging pump. Figure~\ref{fig:example_3_criticality} shows rising criticality across all three models, with anomalies flagged approximately 10 hours before the report. Since the problem was only partially addressed, the criticality does not decrease to zero after the report. ARCANA feature importances are dominated by the upper storage temperature (55\%), while the heat‑circuit‑1 control valve set‑point and primary‑side flow contribute substantially less (15\% and 12\%) to the anomalies detected. As shown in Figure~\ref{fig:example_3_topk}, the models fail to reconstruct the drop in the upper storage temperature, that does not come back up again. After the report the problem is addressed, but not yet resolved. A day later the upper storage temperature decreases again. The identified abnormal opening of the control valve is probably led by the controller trying to achieve the desired upper storage temperature, while the charging pump is not capable of delivering the required flow. These indications can support in the root-cause analysis of the fault.

\subsection{Limitations and future research}
This dataset and baseline results provide a practical starting point for early fault detection in \ac{dhs} under realistic constraints: limited, noisy, and sometimes incomplete training data. The baseline is intended as a reference against which new models can be compared. However, several aspects should be considered in future work.

\begin{enumerate}[I]
    \item Limited samples and labels: The dataset contains relatively few events per fault label. In addition, the substations differ in configuration (e.g., space-heating only vs. combined space-heating and \ac{dhw}, presence of sub‑circuits, etc.) and available signals (see Section~\ref{data}). Because of this sparsity and heterogeneity, we deliberately refrain from reporting per-label or per-configuration performance metrics, and instead focus on aggregated eventwise results. Configuration‑specific statistics can be dominated by small‑sample effects and by the particular mix of faults occurring in each group rather than by the configuration itself. The reported overall detection rate of 60\% and the aggregate accuracy, reliability, and earliness values should therefore be interpreted as benchmark results for this specific operator and dataset, rather than as a universally performance guarantees across all district heating fleets. Future work should extend the dataset with additional operators and faults to assess generalisation across networks, operating contexts and substation types.
    \item Normal-event definition and unlabelled faults: Normal events could be refined to better represent all stable operating regimes across seasons and control modes. An option would be to generate these automatically from the time series, based on the incident reports and maintenance tasks and evaluate the models' ability to recognise normal behaviour on randomly selected events. However, since the time series may contain unlabelled faults, e.g. faults not reported by the customer because they did not influence the comfort level, some visual inspection of the select normal behaviour time series may still be needed. 
    \item Data resolution: Due to the 10‑minute resolution, it is not possible to detect faults such as brief valve oscillations. Some fault types may be under-represented or difficult to confirm (e.g., progressive heat‑exchanger fouling with subtle signatures). A dataset with higher‑frequency data or targeted event logging would therefore be interesting, to test whether these fault types can also be detected beforehand.
    \item Limited training data: As some reports in the dataset only have a small amount of training data available, transfer learning strategies, such as cross‑substation pretraining or domain adaptation between manufacturers, should be researched used to improve early detection results. Physics‑informed methods should also be explored. For example by embedding thermal and hydraulic constraints in the \ac{ae} network or other grey‑box approaches can regularise learning, enforce physical plausibility, and help fault diagnosis. 
    \item Control parameter drift: In addition, incorrect or drifting parametrisation of the control unit (e.g., heating‑curve slope and offset) is a common cause of sub‑optimal operation, contributing to low annual temperature differences and elevated return temperatures \cite{gadd_achieving_2014}. As buildings and customer behaviour change over time, initially adequate settings become misaligned and should be re‑tuned; detecting when control parameter updates are needed is not addressed in this work and should be focus of future research.
    \item Deployment and operator interaction: The present work evaluates models offline; real-time deployment requires context-specific adaptations (e.g., hardware, data infrastructure, maintenance workflows) beyond this study's scope. Future work should bridge detection and action by developing operator decision support tools that translate anomaly patterns and feature attributions into concrete recommendations (inspection, adjustments, repairs), integrated into existing workflows to ensure timely interventions \cite{mansson_faults_2019}.
\end{enumerate}

\section{Conclusions}\label{conclusions}
This work presents a reproducible early fault detection framework for \ac{dhs} by integrating a publicly available, anonymised, service‑report–validated dataset, an evaluation method focused on operational utility, and open‑source baseline implementations with the EnergyFaultDetector, an \ac{ae}-based early fault detection framework for energy systems.

The central outcome of this work is the publicly available, anonymised dataset of 10‑minute operational data from 93 \ac{dhs} from two manufacturers, \emph{M1} and \emph{M2}. It includes annotations about maintenance tasks, customer incident reports, and faults. Annotations include fault descriptions, fault labels and monitoring‑potential ratings, enabling reproducible benchmarking of early fault detection. While limited to a single operator and 10‑minute sampling the dataset provides a realistic foundation for method development and comparison.

We tested three model variants of the EnergyFaultDetector as baseline early fault detection method on this dataset. The conditional \ac{ae} variants outperformed the default \ac{ae} model, however, the day-of-year \ac{ae} also had a higher false-positive rate and slightly reduced normal-event accuracy on \emph{M2}. The conditional \ac{ae} achieves high normal-behaviour accuracy and detects 60\% of the faults pre‑report, with a high reliability and average lead time of about 4 days for detected faults.

To demonstrate practical value, three case studies apply ARCANA for post‑hoc root‑cause analysis: incorrect night‑mode setting for domestic hot water, a failing differential pressure regulator, and a malfunctioning storage charging pump. In each case, ARCANA highlights the features contributing most to the anomalies, consistent with operator findings, supporting diagnosis and actionability.

Future work should focus on expanding the dataset, e.g., cover more fault types, and explore physics‑informed models and transfer learning to improve fault detection. By coupling an open dataset, transparent metrics, and baseline implementations, this benchmark invites consistent comparisons and accelerates practical \ac{fdd} adoption in \ac{dh} O\&M.

\section*{Acknowledgements}
The development of methods presented was funded by the German Federal Ministry for Economic Affairs and Energy (BMWE) through the research project “PreDist” (grant number 03EN3082). The dataset is based on the operational and service data provided by enercity Netz GmbH.

\bibliographystyle{elsarticle-num-names} 
\bibliography{references}



\end{document}